\documentclass{aastex631}

\begin{document}
\newcommand{\cmm}{\ifmmode{\rm cm^{-2}}\else{$\rm cm^{-2}$}\fi}
\newcommand{\cmmm}{\ifmmode{\rm cm^{-3}}\else{$\rm cm^{-3}$}\fi}
\newcommand{\hi}{\ifmmode{\rm HI}\else{H\/{\sc i}}\fi} 
\newcommand{\oh}{\ifmmode{\rm OH}\else{O\/{\sc H}}\fi} 
\newcommand{\glon}{\ifmmode{\ell}\else{$\ell$}\fi} 
\newcommand{\glat}{\ifmmode{b}\else{$b$}\fi}
\newcommand{\vlsr}{\ifmmode{V_\mathrm{LSR}}\else{$V_\mathrm{LSR}$}\fi}
\newcommand{\vgsr}{\ifmmode{V_\mathrm{GSR}}\else{$V_\mathrm{GSR}$}\fi}
\newcommand{\vfor}{\ifmmode{V_\mathrm{for}}\else{$V_\mathrm{for}$}\fi} 
\newcommand{\vdif}{\ifmmode{V_\mathrm{dif}}\else{$V_\mathrm{dif}$}\fi} 
\newcommand{\vtheta}{\ifmmode{V_\mathrm{\theta}}\else{$V_\mathrm{\theta}$}\fi} 
\newcommand{\VR}{\ifmmode{V_\mathrm{R}}\else{$V_\mathrm{R}$}\fi} 
\newcommand{\vz}{\ifmmode{V_\mathrm{z}}\else{$V_\mathrm{z}$}\fi} 
\newcommand{\dg}{\ifmmode{^\circ}\else{$^\circ$}\fi} 
\newcommand {\kms}{\ifmmode{\rm km \, s^{-1}}\else{$\rm km \, s^{-1}$}\fi}
\newcommand{\mo}{\ifmmode{M_\mathrm{\odot}}\else{$M_\mathrm{\odot}$}\fi} 
\newcommand {\Ro}{${\rm R}_\odot$}
\newcommand {\moyr}{\,{\rm M_\odot\,\rm yr}^{-1}}
\newcommand{\nhi}{\ifmmode{N_\mathrm{\hi}}\else{$N_\mathrm{\hi}$}\fi} 
\newcommand{\edt}[1]{{\color{red}#1}}

\shorttitle{Leading Component of the Smith Cloud}
\shortauthors{Lockman et al.}

\title{A Component of the Smith High Velocity Cloud Now Crossing the Galactic Plane}

\correspondingauthor{F. J. Lockman}
\email{jlockman@nrao.edu}

\author[0000-0002-6050-2008]{Felix J.\ Lockman}
\affiliation{Green Bank Observatory, Green Bank, WV 24944, USA}

\author[0000-0002-8109-2642] {Robert A.\ Benjamin}
\affiliation{Dept. of Physics, University of Wisconsin-Whitewater, Whitewater, WI 53190}

\author[0000-0003-4236-2053] {Nicolas Pichette}
\affiliation{Dept. of Physics, Montana State University, Bozeman, MT}

\author[0000-0001-9208-6988] {Christopher Thibodeau}
\altaffiliation{Current Address: 236 Braeden Brooke Dr. San Marcos, TX 78666}
\affiliation{Department of Physics, Astronomy \& Geosciences, Towson University, Towson, MD.}

\begin{abstract}

We have identified a new structure in the Milky Way: a leading component of the Smith high velocity cloud that is now crossing the Galactic plane near longitude $25\arcdeg$.
Using new 21cm \hi\ data from the Green Bank Telescope (GBT) we measured the properties of several dozen clouds that are part of this structure.
Their kinematics is consistent with that of the Smith Cloud with a \vlsr\ exceeding that permitted by circular rotation in their direction.  
Most of the clouds in the Leading Component show evidence that they are interacting  with disk gas allowing the location of the interaction to be estimated. 
The Leading Component crosses the Galactic plane at a distance from the Sun of 9.5 kpc, about 4.5 kpc from the Galactic Center.
Its  \hi\ mass may be as high as $ 10^6$ \mo, comparable to the mass of the neutral component of the Smith Cloud, but only a fraction of this is contained in clouds that are resolved in the GBT data.
Like the Smith Cloud, the Leading Component  appears to be  adding mass and angular momentum to the ISM in the inner Galaxy.
We suggest that the Smith Cloud is not an isolated object, but rather part of a structure that stretches more than $40\arcdeg$ ($\sim 7$ kpc) across the sky, in two pieces separated by a gap of $\sim 1$ kpc.
\end{abstract}

\keywords{ISM: clouds -- ISM: individual objects (Smith Cloud) -- Galaxy: structure -- Galaxy: evolution }

\section{Introduction}
\label{sec:Introduction}
The old puzzle of the Milky Way's high velocity clouds  --  those  aggregations  of neutral and ionized gas whose velocities deviate significantly from that allowed by Galactic rotation -- continues to confound and delight.
The origin of most high velocity clouds (HVCs) is uncertain.
Because the category is defined solely on kinematics \citep[e.g.,][]{Wakker1991}, it can encompass everything from gas stripped from satellite galaxies, accretion from the intergalactic and circumgalactic medium, and material recycled from the Galactic disk \citep{Oort1969, 1997Wakker,  2010Marinacci, Binney2012, Putman2012}.  
HVCs cover a large fraction of the sky \citep{Lockman2002,Lehner2022}.  
{  The large HVC complexes lie in the Galactic halo and are typically found within 10 kpc of the Galactic plane \citep{Putman2017}.}
HVCs have now been detected in other galaxies, e.g., M31 \citep{2004Braun, Thilker2004, Westmeier2008, Wolfe2016}, though not with the sensitivity and  richness of detail available to studies of the Milky Way.

HVCs are important in understanding the evolution of galaxies as they can be manifestations of  both gas accretion and recycling from a galactic fountain \citep{Putman2017, Richter2017, Fraternali2017, LiTonnesen2020}.
{  A number of HVCs have a cometary or head-tail morphology or other indications that they are interacting with an external medium \citep{Putman2011,Barger2020}.
There is also evidence for the interaction of infalling HVCs with neutral gas in the Galactic plane \citep{Lockman2003, McClure-Griffiths2008, Park2016}.
These clouds must have survived their passage through the halo to merge with the Milky Way disk.}

The \object{Smith Cloud} is one of the most prominent high-velocity clouds; it was discovered  even before the identification of HVCs as a class of interstellar object \citep{1963Smith,1963Muller}.
In neutral hydrogen emission the Smith Cloud extends over more than $10^{\circ}$ on the sky with a highly-organized cometary shape.
It has a magnetic field draped  around it \citep{Hill2013, Betti2019}.
The Cloud contains at least $10^6$ \mo\ of \hi\ and probably an equal mass in ionized gas but has no prominent stellar counterpart \citep{Lockman2008,Hill2009, 2015Stark}.  Its  metallicity has been determined through optical emission and UV absorption spectroscopy to be approximately half Solar \citep{Hill2009,2016Fox}, but this may reflect more the medium with  which it has  exchanged material than any initial property of the cloud \citep{Gritton2014, Henley2017, Heitsch2022}.
The distance to the Smith Cloud derived from three different methods is 12.4 kpc, placing it 2.9 kpc below the Galactic Plane  \citep{Putman2003, Wakker2008, Lockman2008}.
The kinematics of the cloud has been analyzed by \citet[hereafter L08] {Lockman2008} who suggest that it is on a course to collide with the Milky Way disk in $\approx 30$ Myr, adding mass and angular momentum to the Galaxy, and possibly triggering a burst of star formation \citep{Alig2018}.
The Smith Cloud has been modelled as the gas remnant of a dwarf galaxy merging with the Milky Way \citep{1998Bland-Hawthorn}, the baryonic component of a dark matter subhalo \citep{Nichols2009, Nichols2014, Tepper-Garcia2018}, and the product of supernove or the Galactic fountain \citep{Sofue2004, Marasco2017}.

This is the first in a series of papers exploring aspects of the Smith Cloud. 
Future papers will report on studies of its structure, interaction with the circumgalactic medium, molecular content, and trajectory.  
Here we describe a major interstellar feature associated with the Smith Cloud: a leading component that is now passing through  the Galactic disk.

HVCs are easily detected at high Galactic latitude where their discrepant velocities stand out most clearly from Galactic disk gas.   
There is no reason to believe, however, that the phenomenon is not widespread, and that a significant fraction of the HVC population has been missed through confusion with gas at permitted velocities \citep{2015Zheng}.
Indeed, distinctions based solely on velocity certainly give a biased view of phenomena in the Galactic halo \citep{Bish2021, Marasco2022}.

\begin{figure*}[htb]
	\centering
	\includegraphics[width=0.75\textwidth]{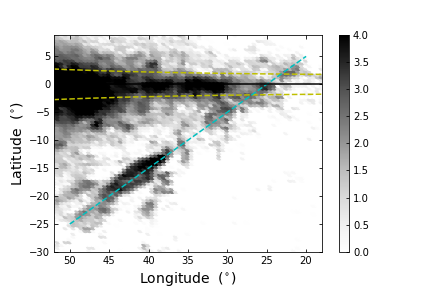}
	\caption{  
		Map of the integrated \hi\  brightness temperature along a slice through the LAB survey that includes only emission between 15 and 35 \kms\ in excess of that allowed by circular Galactic rotation.  
		Dashed yellow lines mark a distance $z = \pm 0.25$ kpc from the plane at the tangent point distance of each longitude.
		Normal thermal and turbulent motions, along with velocity crowding, will produce forbidden-velocity emission in this range whose brightness is expected to increase to higher longitude over areas approximately bound by curves of constant $|z|$. 
		The Smith Cloud is in the lower left between longitudes $40\arcdeg \lesssim \ell \lesssim 50\arcdeg$ centered around latitude $b \approx -15\arcdeg$. 
		This Figure shows that there is also emission that extends  through the Galactic plane along the axis of the Cloud, marked by the diagonal dashed cyan line.  
		We call this the Leading Component.  
		The grey scale is proportional to the square root of the integrated \hi\ brightness temperature.  
		}
		\label{fig:LAB_Smith}
\end{figure*}

In a search for potential HVCs at low Galactic latitude, we constructed a slice through the LAB \hi\ survey \citep{2005Kalberla} in the first Galactic quadrant, integrating over  a velocity interval 15-35 \kms\ greater than expected for circular Galactic rotation for a reasonable Galactic rotation curve.  
{  In the inner Galaxy, the maximum value of $|\vlsr|$ from circular rotation occurs at the tangent point, 
where the distance from the Sun, r, is $R_0 cos(\ell)$, and $R_0$ is the distance to the Galactic Center.
By sampling emission from velocities slightly greater than allowed by circular rotation we are measuring gas in the line wing, which in the absence of non-circular motions should arise from regions close to the tangent point.}
In this instance we used the rotation curve function given by \citet{1993BrandBlitz} with coefficients from \citet{1988Burton}, although as we will show, the basic findings  are not sensitive to details of the adopted rotation curve.
{  
 The results  over  a $40^{\circ}  \times 40^{\circ}$ region are shown in \autoref{fig:LAB_Smith}. } 

{  The \hi\ emission visible in this figure arises from three separate sources.
First,  because of temperature and turbulence there will always be some  emission in the line wings at velocities extending beyond that allowed by Galactic rotation.
The amount of emission in the line wings depends on the velocity dispersion of the gas, but also  on  $|d\vlsr/dr|^{-1}$ -- often called velocity crowding --  which is the degree to which the velocity of gas at different distances from the Sun, r, projects to  a difference in \vlsr\ \citep{Burton1971,Celnik1979}.
The sense  of this geometric effect is that \vlsr\ changes more slowly with distance at high longitudes than at low longitudes, so more emission appears in line wings at a forbidden velocity at 
 $\ell = 50\arcdeg $ than at $\ell = 30\arcdeg$. 
The dashed yellow lines above and below latitude $0\arcdeg$ in \autoref{fig:LAB_Smith} mark a distance $\pm 0.25$ kpc from the plane at the tangent points and  outline expectations for the contribution from line wings and velocity crowding.

A second factor affecting the appearance of \autoref{fig:LAB_Smith} is large-scale streaming motions that arise from, for example,  gas response to spiral arms \citep{Burton1971}.
Although these are easily detected in the data   (e.g., Fig.~8 in \citet{McClure-Griffiths2007} and Figs.~6 and 7 in \citet{Levine2008}), they are not included in rotation curves that describe only the symmetric component.
Large-scale streaming contributes to the emission observed in \autoref{fig:LAB_Smith} around longitude $32\arcdeg$ and $50\arcdeg$.
Note that both turbulence and streaming motions are expected to be approximately symmetric around the Galactic Plane, and that is what is observed.

But \autoref{fig:LAB_Smith} also shows emission not related to Galactic rotation in any sense.  
The Smith HVC dominates latitudes $b \lesssim -10\dg$  between longitudes $35^{\circ} \leq \ell \leq 50^{\circ}$, and there is a component in \hi\ emission almost as extensive as the Smith Cloud that lies along its axis but extends through the Galactic plane, crossing it around longitude $\ell = 25\arcdeg$.
A dashed cyan line connects these two features, which are clearly not symmetric about the Galactic Plane.
The existence of a band of emission at forbidden velocity, aligned with the Smith Cloud,  suggests that both may be part of a larger structure that shares a common  kinematic anomaly. }   
Because the morphology of the Smith Cloud indicates that its space motion is towards lower longitude and more positive latitude (L08), i.e., to the upper right in \autoref{fig:LAB_Smith},  we call the component that extends through the Galactic plane the ``Leading Component".   
To understand  its properties more fully we analyze  new 21cm \hi\ observations from the Green Bank Telescope (GBT) made over a large region for another purpose, and here present those data and the results.

In \autoref{sec:Observations}  the new GBT \hi\ survey is described, as well as an existing study that provides high resolution images of several clouds in the Leading Component.
A catalog of \hi\ clouds found in the Leading Component  in presented in \autoref{sec:Leading Clouds} along with a discussion of their general properties.   
Section 4 discusses evidence that some of these leading clouds are interacting with normally rotating disk gas, allowing derivation of a kinematic distance.  Their overall space velocity is analyzed in  \autoref{sec:SpaceVelocity}, which is followed by a discussion of the implications of the present work.

\section{Observations}
\label{sec:Observations}

\subsection{The GBT \hi\ Survey}

\begin{figure*}[htb]
	\centering
	\includegraphics[width=.95\textwidth]{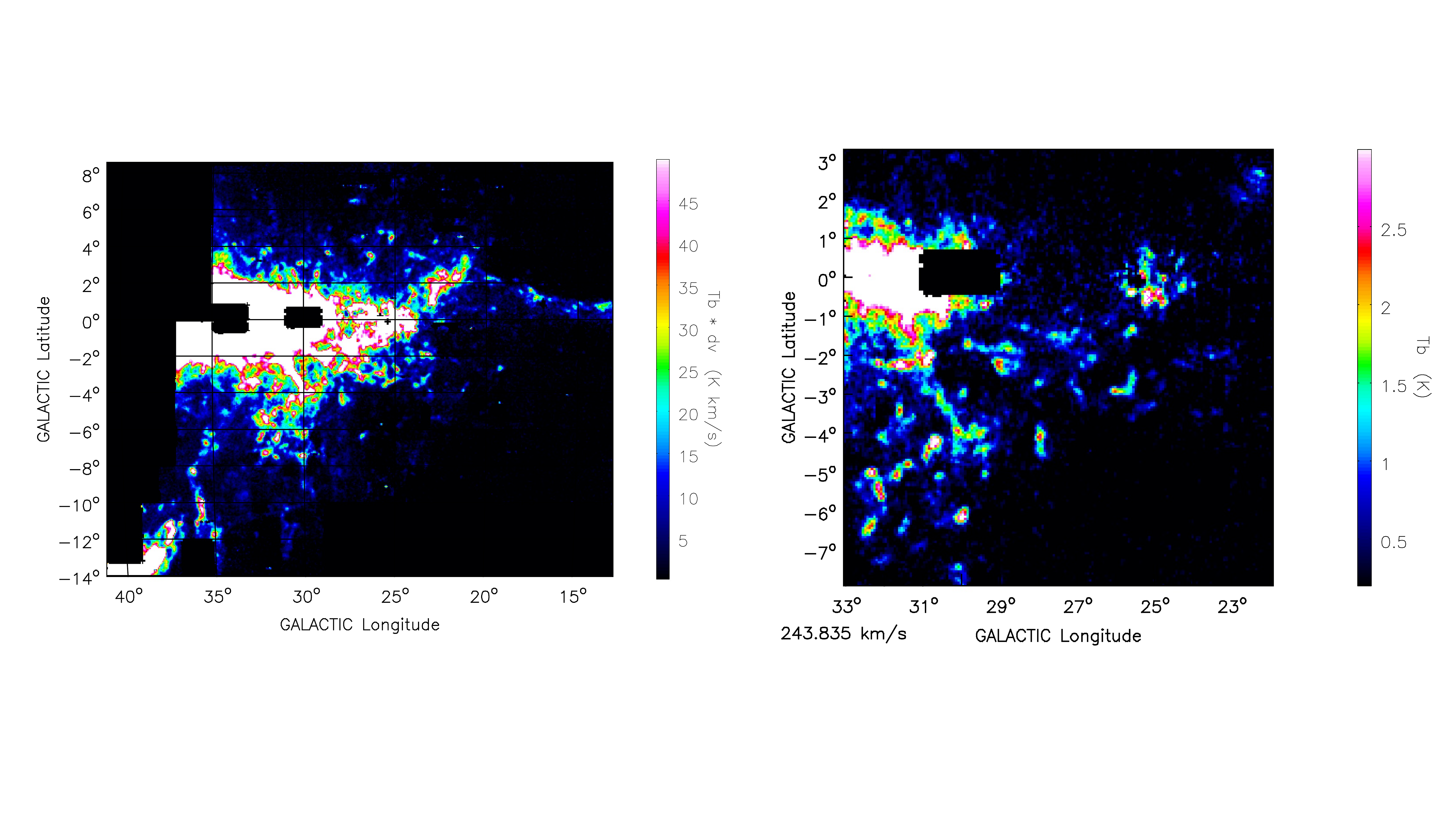}
	\caption{
	Longitude-latitude maps from the new GBT \hi\ survey.
	Coverage is almost complete over $13\arcdeg \leq \ell \leq 35\arcdeg$ and $-10\arcdeg \leq b \leq +10\arcdeg$.
	The new data show the tip of the Smith Cloud in the lower left of the left panel.
	The Leading Component 
	 stands out as a band of emission with the same position angle as the Smith Cloud, extending from $(\ell,b) = (32\arcdeg,-7\arcdeg)$ to $(22\arcdeg,+4\arcdeg)$.
	 Black rectangles mark areas blanked in the GBT data.
	 {\it Left Panel:} The integrated \hi\ emission over  $+226 \leq \vgsr\ \leq  +275 \ \kms$.
	{\it Right Panel:} Channel map at $\vgsr = 243.8\ \kms$ over the central region of the GBT survey.
	Here we see that some of  the Leading Component is resolved into individual clouds. 
		}
		\label{fig:long-lat_VGSR}
\end{figure*}

Measurements of the 21cm line of \hi\ were made with the Green Bank Telescope (GBT) using the standard L band receiver and the GBT spectrometer \citep{2009Prestage}.  
Data were taken "on-the-fly" using in-band frequency switching and were calibrated as described in \citet{2011Boothroyd}.
A low order polynomial was fit to emission-free regions of the spectra  after they were gridded into a cube with a channel spacing of 0.80 \kms\ over a velocity range of $-200 \leq {\rm V_{LSR}} \leq +275$ \kms.
The data cover $13^{\circ} \leq \ell \ \leq 35^{\circ}$ over a latitude range $\pm10^{\circ}$ with {  a partial extension to latitude $-14^{\circ}$ at $ \ell \geq 30^{\circ}$} to include the tip of the Smith Cloud.
The effective angular resolution is $10\arcmin$. 
The median rms noise in brightness temperature for a 1 \kms\ channel is 0.12 K.  
The noise varies somewhat across the mapped area, being greatest within $2^{\circ}$ of the Galactic plane where there was often a  significant increase in the system temperature from continuum emission.  
A small fraction of the mapped area had to be blanked over regions of strong continuum emission or because of occasional radio frequency interference.

For study of the Leading Component, the new data have a distinct advantage over the premier existing \hi\ survey, HI4PI \citep{2016HI4PI}.  
Emission from the leading component is not particularly faint, so although HI4PI has about one-third lower noise level than the new GBT data, noise is not a critical factor.  
The $10\arcmin$ angular resolution of the GBT data compared with the $16\farcm2$ of HI4PI  is, however, important in sorting out confusion between the leading component and other \hi\ emission.  
Virtually all of the Leading Component lies at J2000 declination $\delta < -5\dg$, which is outside the declination limit of any survey using the Arecibo telescope.

The new GBT survey is shown in \autoref{fig:long-lat_VGSR} as emission integrated over forbidden velocities (left), i.e., those velocities not permitted by circular Galactic rotation,  and as a map of emission in a single velocity channel (right).
The left panel shows the tip of the Smith Cloud at the lower left and the Leading Component in the center of the Figure. 
The velocity with respect to the Local Standard of Rest, \vlsr, is not appropriate for the study of objects that cover a large angle on the sky, as the changing projection of the LSR across the structure becomes significant.  A more suitable measure is the velocity with respect to the Galactic Standard of Rest  defined as $\vgsr = \vlsr + |V_0|  sin(\ell) cos(b)$, where V$_0$ is the circular velocity of Galactic Rotation at the Sun's distance from the Galactic Center, ${\rm R_0}$.  
Throughout this paper we use ${\rm R_0} = 8.1$ kpc  and ${\rm V_0 = 230}\  \kms$, consistent with the results of recent studies \citep{2018Gravity,2019Eilers}.
The right panel of \autoref{fig:long-lat_VGSR} is a channel map over the central part of the new GBT survey  at a constant \vgsr\ of  243.8 \kms, showing that  the Leading Component contains discrete clouds.

\subsection{High Angular Resolution Measurements}
\label{sec:Pidopryhora}

\citet[hereafter P2015]{Pidopryhora2015} combined GBT and VLA data  to produce \hi\ maps of 10 interstellar clouds at $\sim 1\arcmin$ angular resolution,  corresponding to a linear resolution of 3 pc at a distance of 10 kpc. 
The clouds cover a range of longitude ($16\fdg0 \leq \ell \ \leq \  44\fdg8$) and  latitude ($-8\fdg7 \ \leq  b \leq  +4\fdg5$).
The sample was selected to have little confusion with other neutral \hi, so these  clouds are somewhat outside of the Galactic plane (median $|b| = 4\fdg9)$ and have a velocity beyond the terminal velocity ($\vlsr \geq \ V_t$), where the terminal velocity is the maximum permitted by Galactic rotation in their direction.  
Three of the P2015 clouds turn out to be clouds in the Leading Component.
Although the noise in the P2015 data is about three times worse than the GBT data (approximately 0.36 K in a 1 \kms\ channel) its much higher angular resolution reveals important structure within the clouds.

\section{Leading Clouds}
\label{sec:Leading Clouds}

\begin{figure*}[htb]
	\centering
	\includegraphics[width=1.0\textwidth]{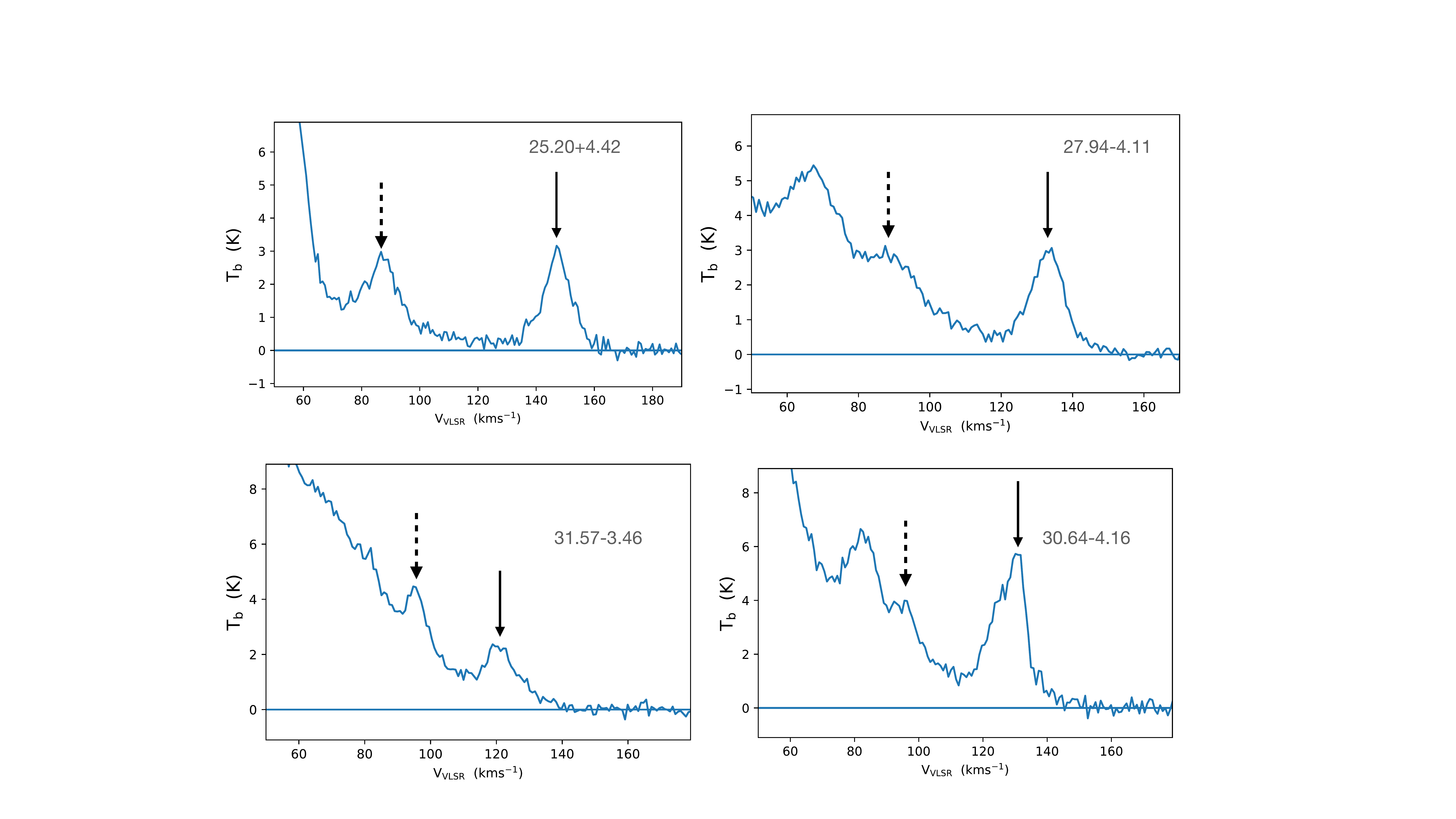}
	\caption{
		GBT \hi\ spectra showing examples of the clouds detected in this survey associated with the Leading Component of the Smith Cloud {  identified by their $\ell,b$ coordinates.}  
		The leading cloud is always the highest velocity spectral feature, marked with the solid arrow.
		All these clouds appear to be interacting with a component of disk gas at a lower (permitted) velocity, marked with a dashed arrow.  
		This is discussed in more detail in \autoref{sec:interaction}.
		}
		\label{fig:LeadingCloud_spectra}
\end{figure*}

\begin{figure*}[htb]
	\centering
	\includegraphics[width=0.8\textwidth]{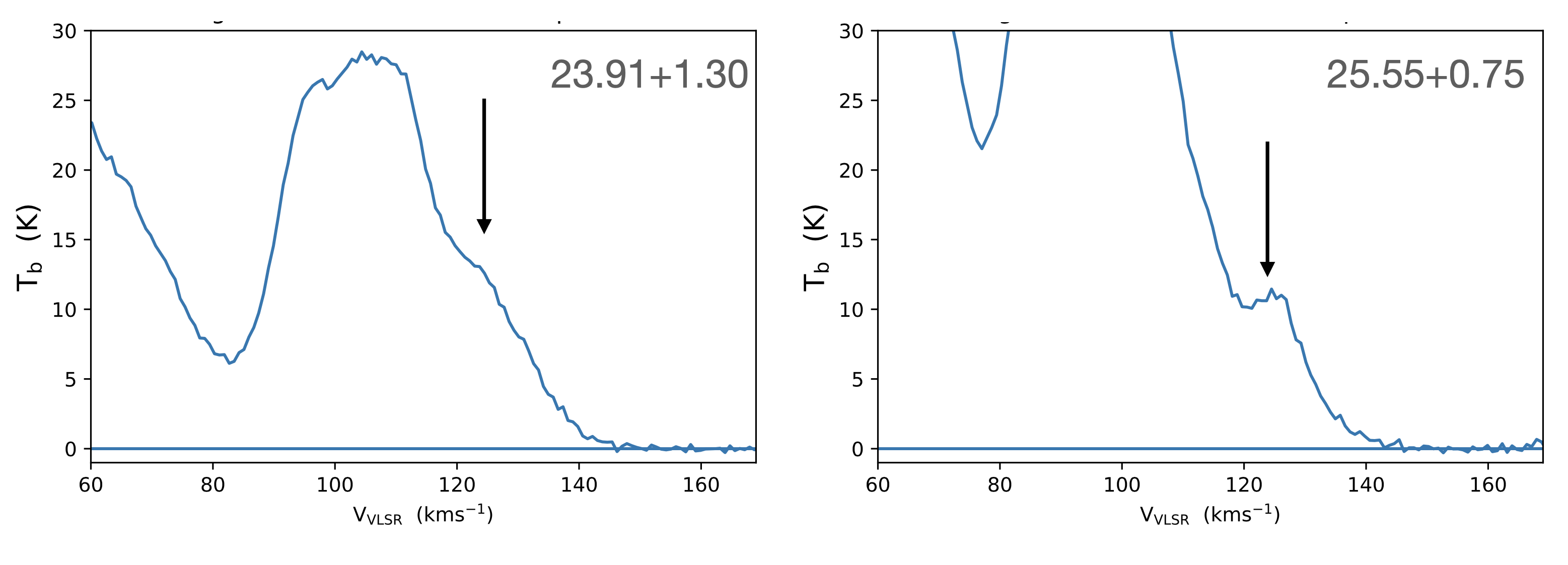}
	\caption{
		GBT \hi\ spectra of Leading Component clouds  that are confused with lower velocity emission (arrows). 
		{  The longitude and latitude is indicated in each panel. }
		These spectral features are distinct clouds in channel maps and have a \vlsr\ exceeding that permitted by Galactic rotation, but cannot be separated from lower velocity emission and, like others with similar properties, are therefore not included in  Table 1.
		}
		\label{fig:Confused_spectra}
\end{figure*}

\subsection{Cloud Properties}
\label{properties}

At the angular resolution of the GBT data the Leading Component contains a number of individual clouds many of which are isolated in $\ell$, b, and velocity, so that their properties can be measured accurately.  
The right panel of \autoref{fig:long-lat_VGSR} shows a region of the survey centered near $(\ell,b) = (27\arcdeg, -2\arcdeg)$ in which a number of individual clouds can be identified.  
The clouds lie in a swath that begins a few degrees above the top of the Smith Cloud around $(\ell,b) = (35\arcdeg, -8\arcdeg)$  and extends at least to $b > +4\dg$ at $\ell < 24\dg$. 
Discrete clouds associated with the leading component were identified using three criteria:
 1) a location on or near the main region of the leading component; 2) a \vlsr\ that was close to or beyond that permitted by circular Galactic rotation in their direction; and, 3) for most of the clouds, evidence that they were interacting with gas at a lower, permitted velocity.
Spectra towards 4 of the leading clouds are shown in \autoref{fig:LeadingCloud_spectra}.  
They are detected well above the noise in the GBT data.  

There are also cloud-like structures visible in channel maps whose spectral components  cannot be separated from lower velocity emission.  
This is especially common close to the Galactic plane.
Examples are shown in \autoref{fig:Confused_spectra}.
These are not included in Table 1.

Clouds were identified in channel maps and a Gaussian was fit to the spectrum at the location of the peak \nhi.
A 2-d Gaussian model was used to derive an angular size and position angle.  
\hi\ masses were measured by integrating over an area of the channel map encompassing  the measurable extent of the cloud at the velocity of the \hi\  peak.
Background \hi\ emission was estimated from measurements around the edge of the cloud and subtracted. 
The result was then multiplied  by the FWHM of the line at the brightest location to obtain a line power integrated over area and velocity.  
Some clouds were so confused with other emission that it was not possible to  measure their size.

Properties of 38 clouds are given in Table 1.  
We emphasize that the sample of clouds presented here does not completely account for all the clouds or all the \hi\ mass in the leading component;  it  represents only the more easily-identifiable and less-confused clouds in the GBT data. 
The longitude and latitude of the peak \hi\ brightness is in the first two columns, and quantities in cols. 3 through 6 refer to that location.  
Errors in line parameters were derived from the Gaussian fit to the spectrum. 
The median peak $T_b$ of the clouds is 2.5 K which is a detection at the $20\sigma$ level above the noise.  
The faintest cloud is detected at $6.5\sigma$.
The angular size  in col.~7 is not corrected for beam smearing.  
The position angle is measured counter-clockwise with reference to the Galactic North Pole such that a cloud elongated parallel to the Galactic plane would have PA $= 90\arcdeg$.  
For the purposes of calculating general cloud properties we assume that they are all at a distance $d_{10} = 10$ kpc.
This is comparable to the distance of the Smith Cloud, 12.4 kpc (L08), and results can be  scaled easily.
The distance of the Leading Component is discussed further in \autoref{sec:interaction}.

\begin{figure*}[htb]
	\centering
	\gridline{\fig{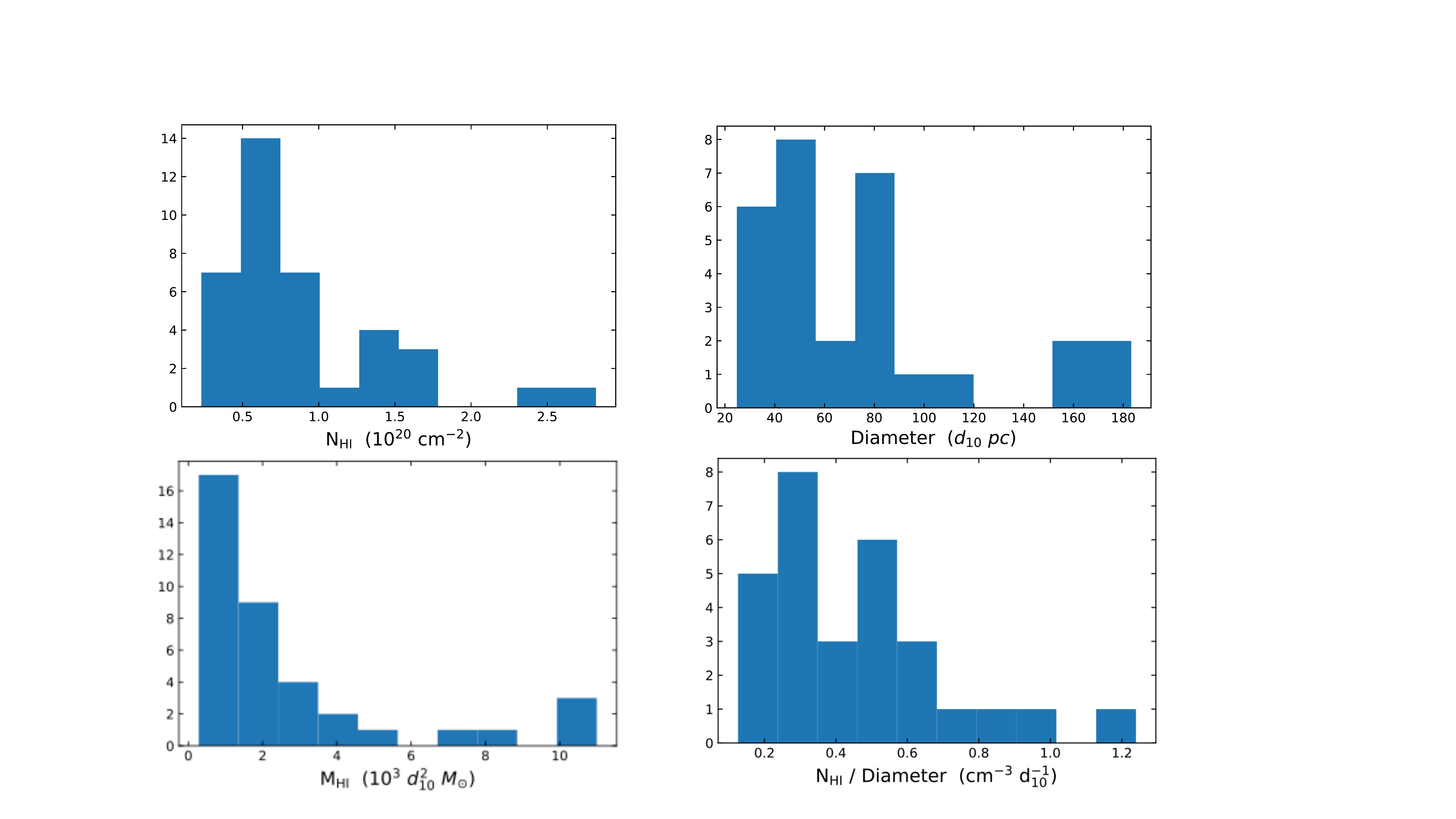}{0.97\textwidth}{}}
    \gridline{\fig{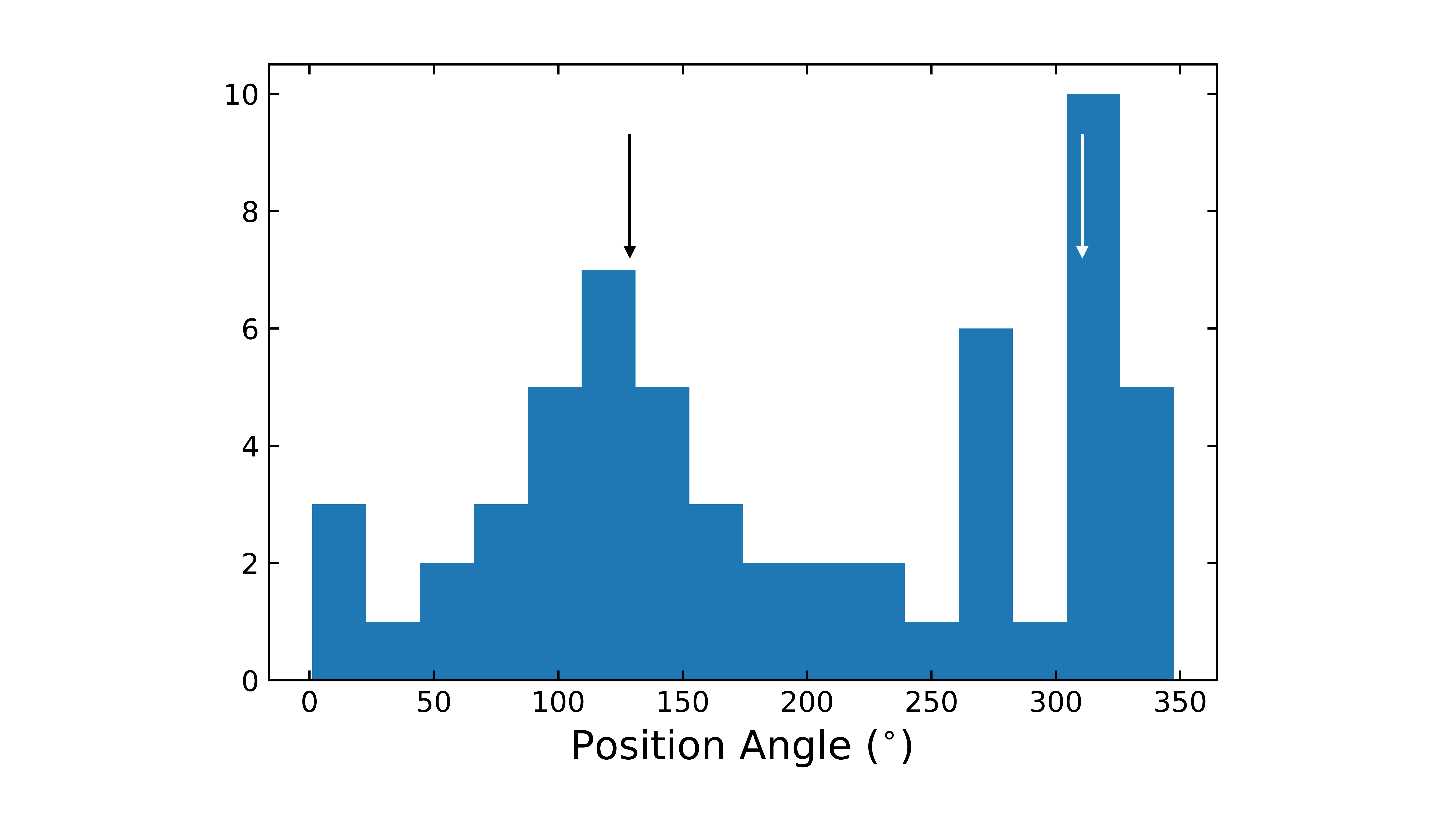}{0.48\textwidth}{}}
	\caption{
	Properties of the leading clouds.
		{\it Top left:} \nhi\ measured at the position of  peak line brightness assuming optically thin emission. 
		{\it Top right:} Cloud diameter deconvolved for beam smearing  assuming a 10 kpc distance.
		{\it Center left:} \hi\ mass of the leading clouds assuming a distance of 10 kpc.  
		The median $M_{\hi} = 1.6 \times 10^3 \ \ (d_{10}^2) \ \mo $.
		{\it Center right:} Estimate of the average volume density in the clouds from the ratio of the peak column density to the diameter.  The diameter is corrected for beam convolution and scaled to a distance $d_{10} = 10$ kpc.  The median $n_{HI} = 0.43 \ \cmmm$.
		{\it Bottom:}
	Distribution of position angles for the leading clouds.  
	Each cloud is displayed twice, once at its PA and once at its PA+180$^{\circ}$.
	The two vertical arrows mark the PA of the Smith Cloud, $130^{\circ}$.
	The data suggest that many individual clouds  share the same elongation as the  Smith Cloud.
		}
		\label{fig:histograms}
\end{figure*}

Properties of the leading cloud population are summarized in \autoref{tab:properties}. 

The measured angular size of the clouds, $\Theta_{obs}$, was corrected for beam convolution by subtracting the $10\arcmin$ resolution of the data cube in quadrature from the measured $\Theta_{maj}$ and $\Theta_{min}$.  
Cloud diameters were derived from the square root of the product of the major and minor deconvolved angular sizes,  at a distance of 10 kpc.
They are shown in the upper right panel of \autoref{fig:histograms}.

The cloud \hi\ mass distribution is shown in \autoref{fig:histograms} for an assumed distance of 10 kpc.
The sum of the \hi\  mass of the tabulated clouds is $1.1 \times 10^{5}$ \mo.  
The \hi\ mass in the entire leading component was estimated by summing over a large region encompassing the leading component, and integrating over all $\vgsr \geq 226$ \kms.
The result, $8 \times 10^{5}$ \mo\ for a distance of 10 kpc, has a large uncertainly, but implies that the tabulated clouds contain only a small fraction ($\sim15\%$) of the total \hi\ associated with the system.

A lower limit to the volume density of the clouds can be derived from the ratio of the peak column density, \nhi, to the cloud diameter.  
This is a lower limit because the observed \nhi\ is certainly reduced to some extent by beam convolution, while this effect was taken out, at least approximately, in calculating the diameter.

{  The distribution of cloud position angles is shown in the bottom panel of \autoref{fig:histograms}, where to avoid artificial discontinuities at $180^{\circ}$ each cloud is counted twice, once at its PA and once at its PA+180$^{\circ}$.  
The distribution peaks around the PA of the Smith Cloud, which is also the PA of the Leading Component, $130^{\circ}$, suggesting that some individual clouds are being shaped by the same interaction with their environment as the Smith Cloud itself. 
}

\subsection{Comparison with higher angular resolution observations}
\label{sec:P2015comp}

Three of the leading clouds from Table 1 were observed  in the P2015 survey and we can compare properties derived from GBT data with those from the higher angular resolution data.
The results are given in \autoref{tab:resolution_comp}, where  P2015 diameters have been scaled to a distance of 10 kpc. 
Velocities are in good agreement, while the higher angular resolution of P2015 gives larger peak values of \nhi\ by a factor of 2 - 3.
The higher resolution data often reveal cloud cores with a diameter less than one-third that given by the GBT data.
The increase in peak \nhi\ and smaller cloud core size combine to increase the central cloud density by factors of more than six.
It is interesting that these clouds are elongated along the direction of the Leading Component axis, with position angles clustered around the $130\arcdeg$ PA of the Smith Cloud.

\subsection{Leading Cloud Kinematics}

\begin{figure*}[htb]
    \centering
    \includegraphics[width=0.7\textwidth]{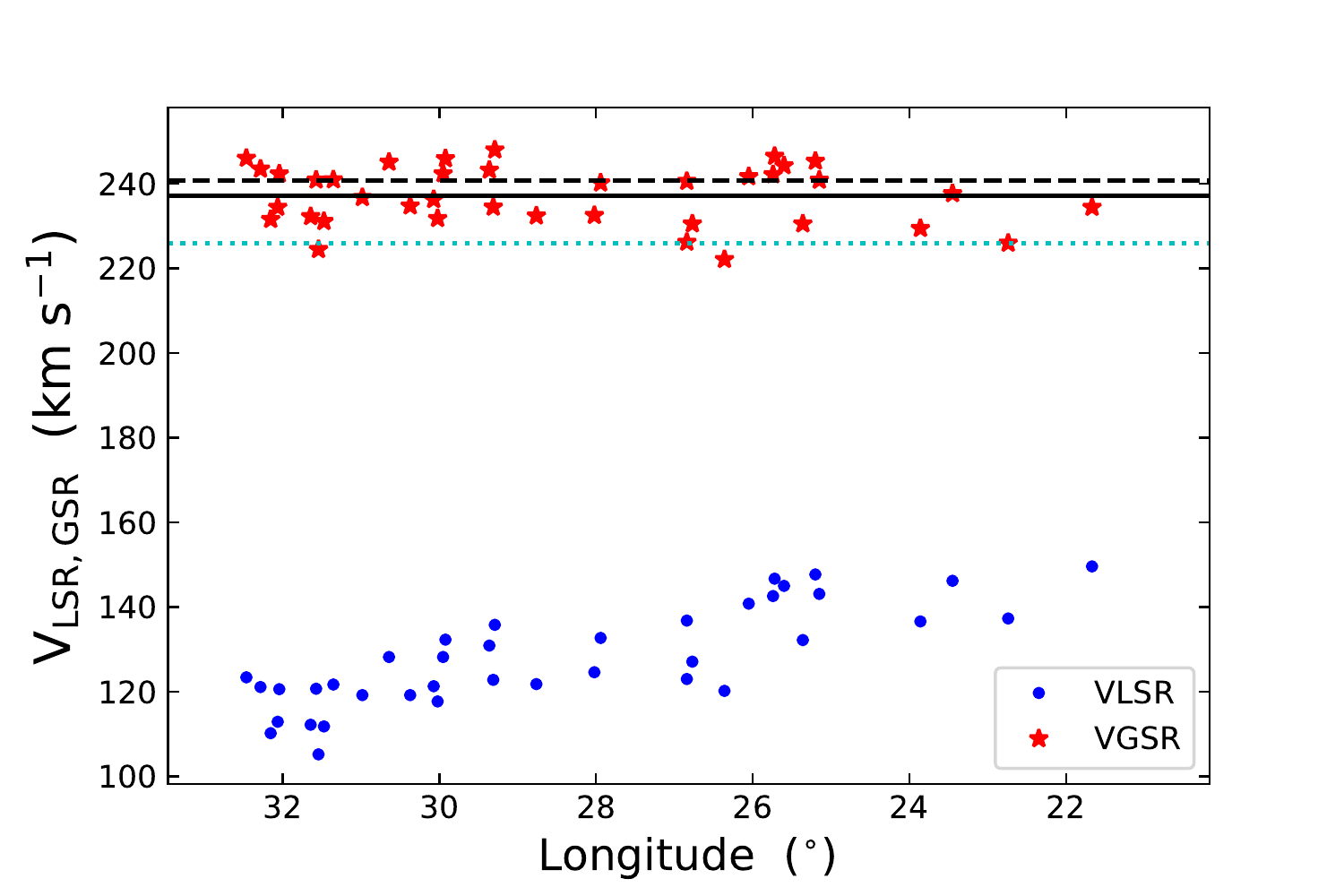}
    \hspace{0.02\textwidth}
    \caption{ 
    {  The leading clouds in position and velocity. 
    The \vlsr\ (blue dots below) and \vgsr\ (red stars above) of the leading cloud sample is plotted vs. longitude.}
    The solid horizontal line is the median \vgsr\ of the clouds, 237 \kms, while the dashed line is the \vgsr\ of the Smith Cloud, 241 \kms.
    There is no strong trend in \vgsr\ with longitude implying that the entire complex is consistent with having a single velocity vector projected to different values of \vlsr\ at different locations.
    The dotted line shows the lower limit to the \vgsr\ used in integrating over velocity to create the moment-zero map in the left panel of  \autoref{fig:long-lat_VGSR}.
    }
    \label{fig:Long_VLSR_VGSR}
\end{figure*}

{  Values of \vlsr\ and \vgsr\ for the clouds are shown  in a position-velocity diagram} in  \autoref{fig:Long_VLSR_VGSR}.  
The solid line is at  the median \vgsr\ of the sample, 237  \kms, while 
the mean \vgsr\ is identical at $237.1\pm6.8$ \kms ($1\sigma)$. 
The \vgsr\ of the central region of the Smith Cloud, marked by the dashed line, is  essentially identical to that of the leading clouds at 241 \kms.
The dotted line at 226 \kms\ marks the lower velocity limit of the integrated  emission displayed in the left panel of \autoref{fig:long-lat_VGSR}, which was used to estimate the \hi\ mass of the entire Leading Component system.

{ 
As we will discuss in \autoref{sec:SpaceVelocity}, the quantity \vgsr\ is directly related to the total space velocity of an object.  
The combination of similar \vgsr\ and an apparent association on the sky strongly supports a physical connection between the Leading Component and the Smith Cloud. }

From a  rotation curve which gives $V_{\theta}(R)$, a terminal (maximum) velocity can be calculated for every longitude $|\ell| < 90\arcdeg$:
$Vt \equiv cos(b) \ [V_{\theta}(R_0) sin(\ell) - V_{\theta}(R_t)]$, where $R_t =  R_0 \ sin(\ell).$
Each cloud can then be assigned a "forbidden" velocity, $\vfor \equiv \vlsr - V_t$, which gives the amount by which the cloud's \vlsr\ is in excess of that allowed by circular Galactic rotation {\it at any location along that line of sight}.  

We use two rotation curves:  
 a flat rotation curve with $V_{\theta}(R) \equiv V_0 = 230 \  \kms$, which is very close to the curve measured by \citet{2019Eilers} at $R \geq 5.27$ kpc, 
and  a "Universal" rotation curve \citep{1996Persic} with coefficients from \citet{2019Mroz}, that we call the PM model.
The PM model differs from that of \citet{Reid2014} by only a few percent over the region of interest.

\begin{figure}
    \centering
    \includegraphics[width=0.95\textwidth]{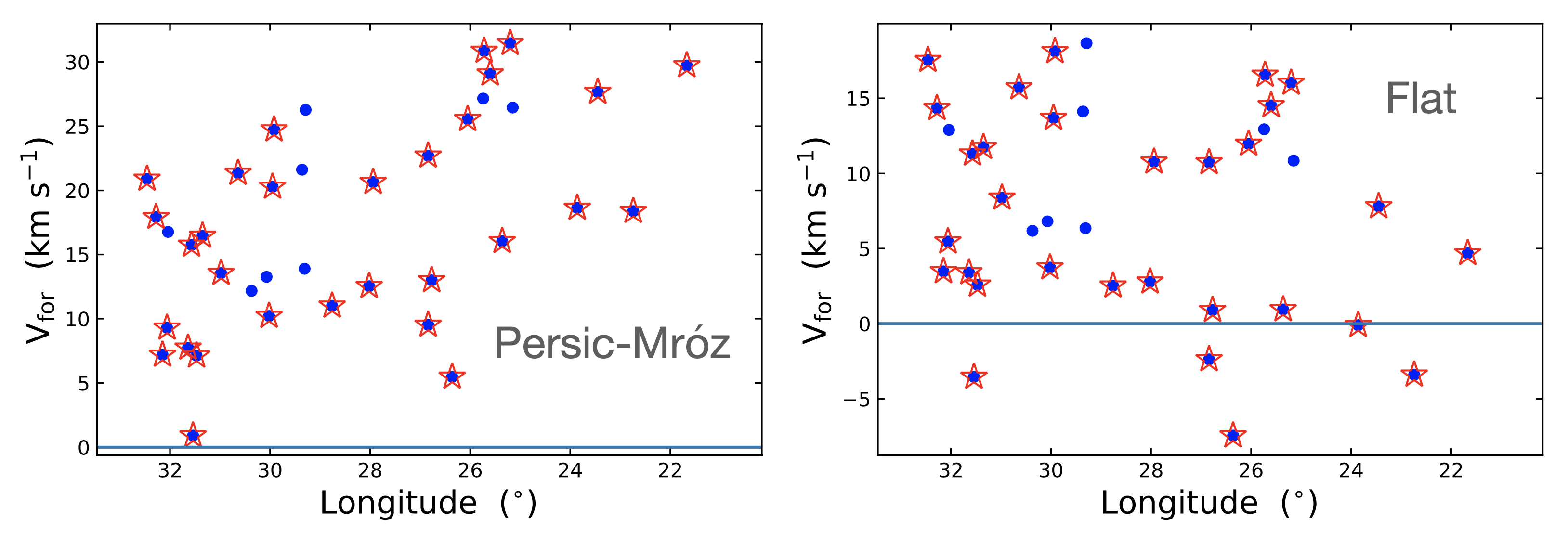}
    \hspace{0.02\textwidth}
    \caption{ {  The "forbidden velocity" of the cloud sample, \vfor, defined as the difference between a cloud's \vlsr\ and the maximum permitted by circular Galactic rotation in their direction  (blue dots).}
    This is a lower limit to the peculiar line-of-sight velocity of a cloud.  
    A more realistic estimate of a cloud's deviation from circular rotation is given by the quantity \vdif, discussed in \autoref{sec:interaction}.
    If the clouds show signs of interaction with lower velocity gas the points are overplotted  with red stars.
   {\it Left panel:} Forbidden velocities evaluated for the Persic "Universal" rotation curve \citep{1996Persic} with coefficients from \citet{2019Mroz}.
   {\it Right Panel:}  Forbidden velocities for a flat rotation curve at a circular velocity of 230 \kms.
    While there are  5 clouds with $ \vfor \lesssim 0 $  for a flat rotation curve,  values for the more realistic Persic curve, and  evidence for interaction, suggests that they have anomalous velocities compared to their location in the Galaxy and are properly included in the leading cloud sample.
    }
    \label{fig:Long_Vfor}
\end{figure}

\autoref{fig:Long_Vfor} shows the "forbidden velocities" \vfor\ for the cloud sample plotted against longitude from the PM rotation curve on the left and the flat rotation curve on the right.
The median value of \vfor\ is 17.3 and 8.1 \kms, respectively.
Forbidden velocities are larger in the PM model because, unlike the flat model, the  PM rotational velocities decrease toward the Galactic Center reducing the maximum permitted \vlsr\ at lower longitudes.
When analyzed with a flat rotation curve there are five clouds with $\vfor \lesssim 0 \ \kms$,  but these all show signs of interaction with lower-velocity gas implying that they have a high non-circular velocity component (discussed in \autoref{sec:interaction}).  
In \autoref{fig:Long_Vfor} clouds with evidence of interaction are marked with a red star.

\section{Interaction with Disk Gas}
\label{sec:interaction}

\begin{figure*}[htb]
    \centering
    \includegraphics[width=1.0\textwidth]{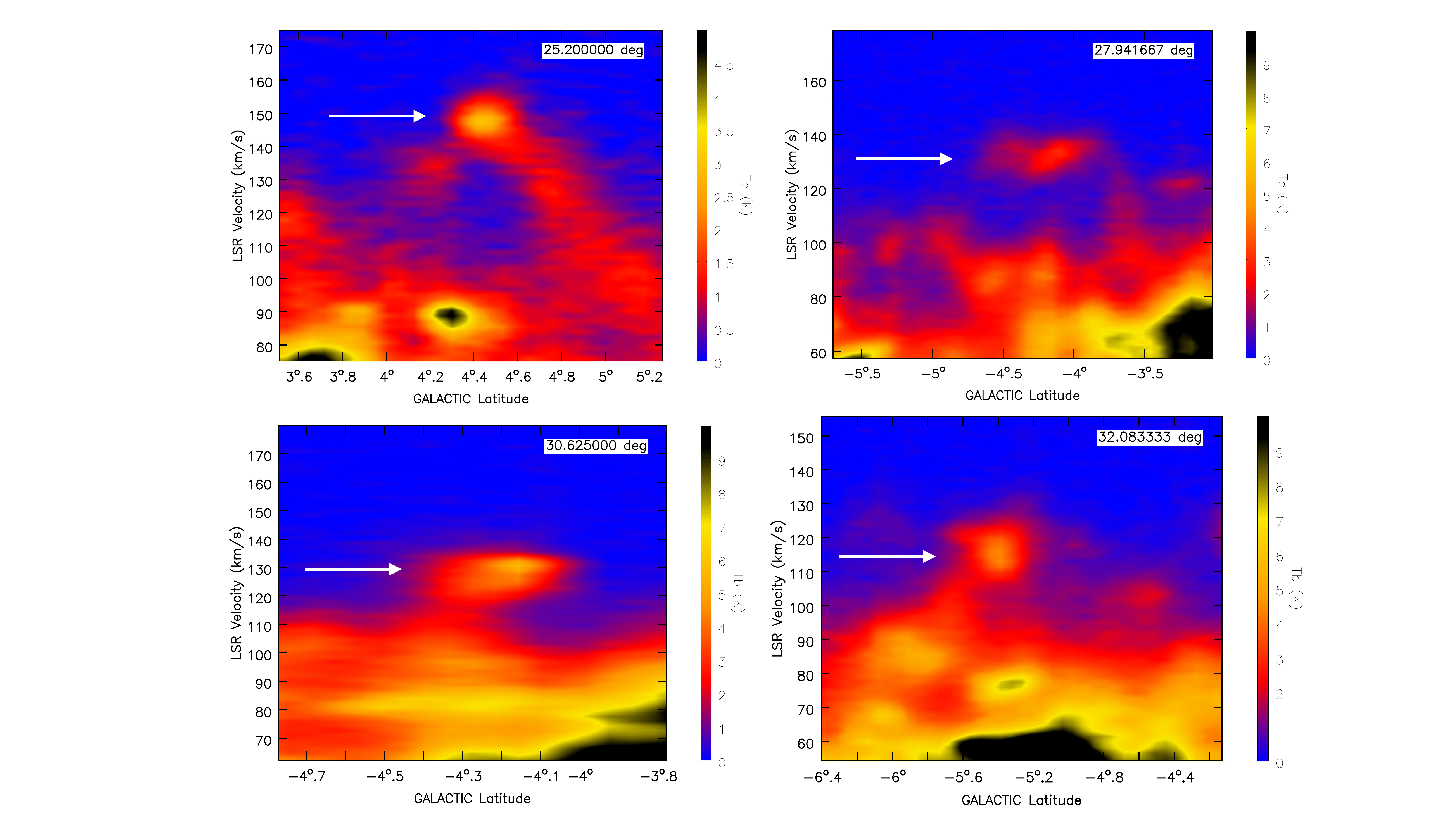}
    \hspace{0.02\textwidth}
   
    \caption{Examples of leading clouds showing their  interaction with the Milky Way disk gas.  
    In these latitude-\vlsr\ plots the leading cloud is always the feature at the highest velocity, marked with an arrow.  
    There are "bridges" to lower velocity \hi\ emission connecting these clouds to gas at a velocity consistent with normal Galactic rotation.
    This is often accompanied by an enhancement in the lower velocity emission.  
    We interpret these figures as showing that these leading clouds are interacting with normal disk gas, with some cloud material being stripped away and decellerating to the lower, permitted velocity.
    }
    \label{fig:BV_maps}
\end{figure*}

\begin{figure*}
    \centering
    \includegraphics[width=0.5\textwidth]{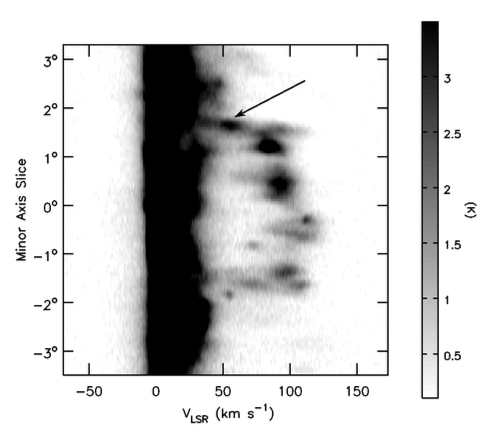}
    \hspace{0.02\textwidth}
   
    \caption{From \citet{Lockman2008}, a  position-velocity cut through the minor axis of the Smith Cloud, showing the same transition at the cloud edge between anomalous velocity gas and normally-rotating disk gas as is seen at the edges of  the leading clouds of \autoref{fig:BV_maps}.
    The arrow marks a small cloud that we interpret as having been detached from the Smith Cloud and decellerated by its interaction with the circumgalactic medium of the Milky Way.
    }
    \label{fig:BV_Smith}
\end{figure*}

 Most of the clouds appear to be interacting with gas at lower $\vlsr$, i.e., with regularly rotating material in the Milky Way disc. 
In some cases the spectra show continuous emission between the velocity of the leading cloud and a lower-velocity component,  overlapping in position with the cloud.
The spectra in \autoref{fig:LeadingCloud_spectra} show four examples, where the lower velocity gas is marked with a dashed arrow.
In many cases the connection is especially convincing, as illustrated in \autoref{fig:BV_maps}.  
Here, in a display of latitude vs. \vlsr, we can see one or more edges of leading clouds forming a continuous bridge of \hi\ emission to lower velocities, consistent with the outer regions of the clouds being stripped and decelerated.
We compare \autoref{fig:BV_maps} to \autoref{fig:BV_Smith}
(from L08), which is a velocity-position cut through the minor axis of the Smith Cloud showing the same kinematic bridge between the anomalous velocity of the Smith Cloud and gas in the Milky Way disk.

\begin{figure*}
    \centering
    \includegraphics[width=1.0\textwidth]{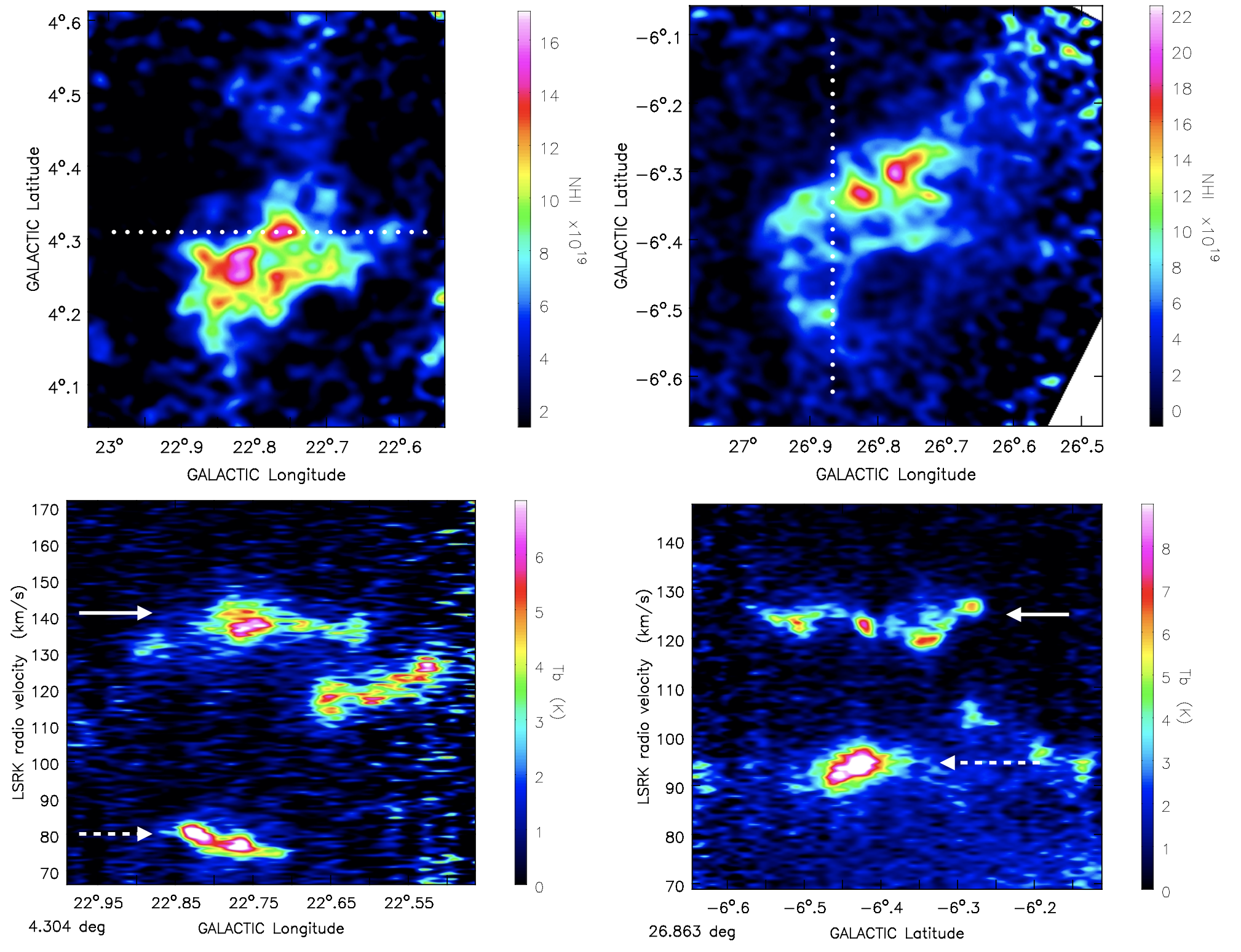}
    \caption{
    {\it Top panels:}
      High angular resolution images of the total \nhi\ from the P2015 data for two of the leading clouds observed in that survey: 22.74+4.30 (left) and 26.84-6.38 (right).  
      {\it Bottom:}	 Cuts through the clouds in longitude-velocity (22.74+4.30, left) and latitude-velocity (26.84-6.38, right) along the directions given by dotted lines in the upper panels. 
      The leading clouds are marked with a solid arrow and a dashed arrow identifies the Milky Way disk gas with which they are interacting.
      Unlike the examples in \autoref{fig:BV_maps}, these data do not show kinematic bridges between the clouds and Milky Way gas.
      Instead, we derive evidence of interaction from the high degree of spatial coincidence between the gas at forbidden and permitted velocities, coincidence 
       which is not  seen in P2015 clouds except for those that are part of the Leading Component.
      The velocity of the interacting gas is used to derive a kinematic distance to the interaction.
    }
    \label{fig:P2015_clouds}
\end{figure*}

The high-resolution P2015 data give additional information on the interaction for two leading clouds: G22.74+4.30 and G26.84-6.38 (we refer to these clouds by their GBT positions in Table 1, which deviate somewhat from the location of the brightest \nhi\  in the P2015 data). 
A map of total \nhi\ and a position-velocity cut through each cloud is shown in \autoref{fig:P2015_clouds}.
Unlike the clouds in \autoref{fig:BV_maps}, these clouds do not exhibit kinematic bridges of material that join the cloud  with lower-velocity gas, but 
in both cases there is a bright \hi\ emission component centered approximately at the position of the leading cloud, but at a much lower velocity. 
The interaction between cloud and disk gas for G25.20+4.42 is shown more clearly in the GBT data of \autoref{fig:BV_maps} than in the P2015 data because the P2015 data do not have the sensitivity to detect the extended bridge of emission that connects this cloud kinematically to the disk.

It is interesting and perhaps significant that of the 10 clouds in the P2015 sample, only those three associated with the Leading Component show unambiguous evidence of interaction with gas at lower velocities.  

\begin{figure*}
    \centering
    \includegraphics[width=0.5\textwidth]{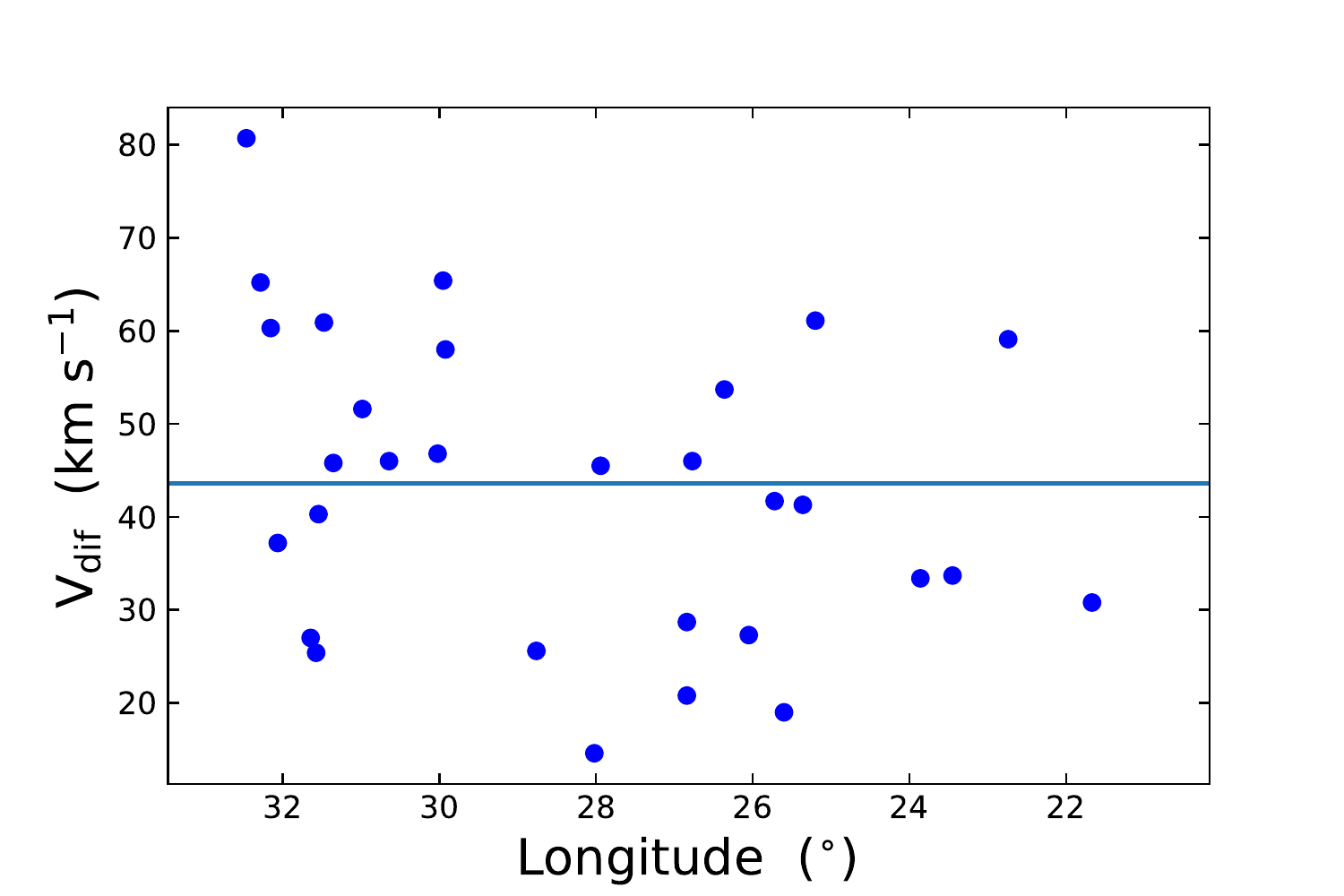}
    \caption{
    {  The line-of-sight} velocity difference \vdif\ between the leading clouds and interacting gas,  plotted against longitude.   
    The median value is 44 \kms, and this is the best estimate of the line-of-sight  deviation of the leading clouds from circular rotation at their location.
    }
    \label{fig:Long_interact}
\end{figure*}

 In Table 1 the \vlsr\ of gas interacting with leading clouds is given as $V_{MW}$.
 This quantity is blank for clouds without identifiable interaction. 
The quantity \vdif, which is the difference in line-of-sight velocity between the cloud and the disk gas that it is encountering, $\vdif \equiv \vlsr - V_{MW}$, is shown in  \autoref{fig:Long_interact}.

The evidence for interaction provides two important pieces of information: 
\begin{enumerate}
  \item the leading clouds are in fact kinematically anomalous by a significant amount; 
  \item the velocity of the interacting gas might be used to derive a distance to the interaction.  
\end{enumerate}

\subsection{Location of the Interaction}
\label{sec:location}

\begin{figure}
\gridline{\fig{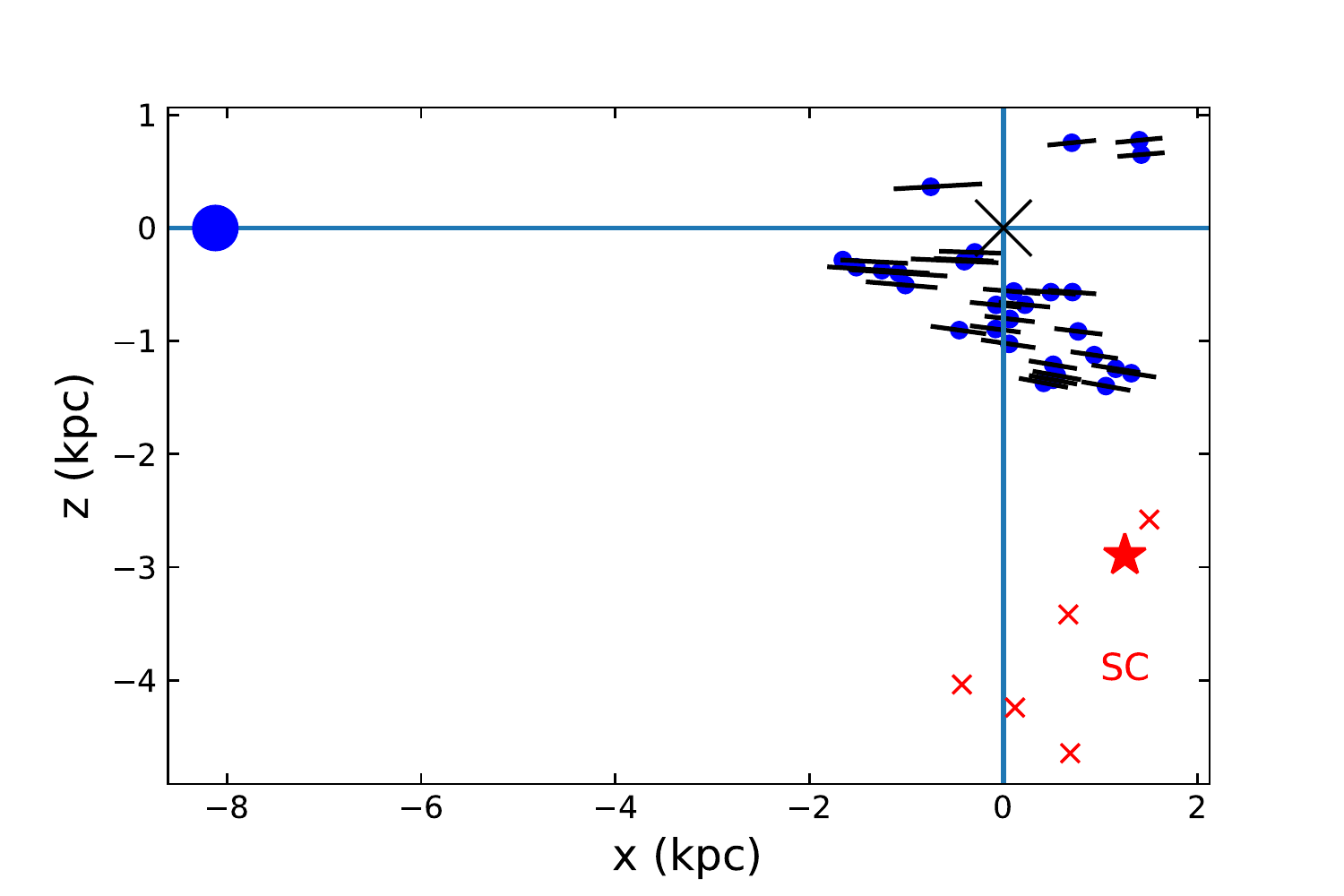}{0.45\textwidth}{(a)}
          \fig{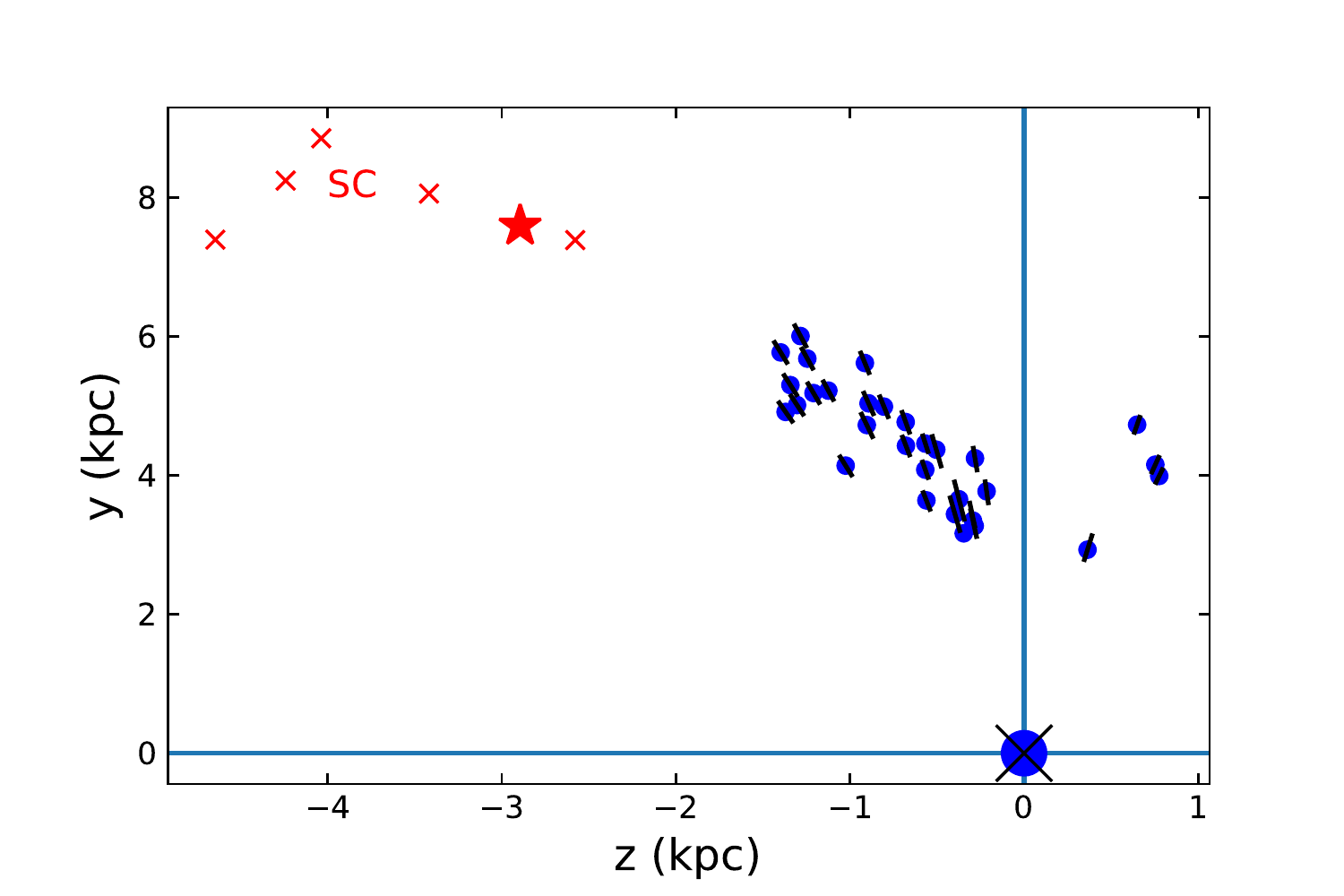}{0.45\textwidth}{(b)}}
\gridline{\fig{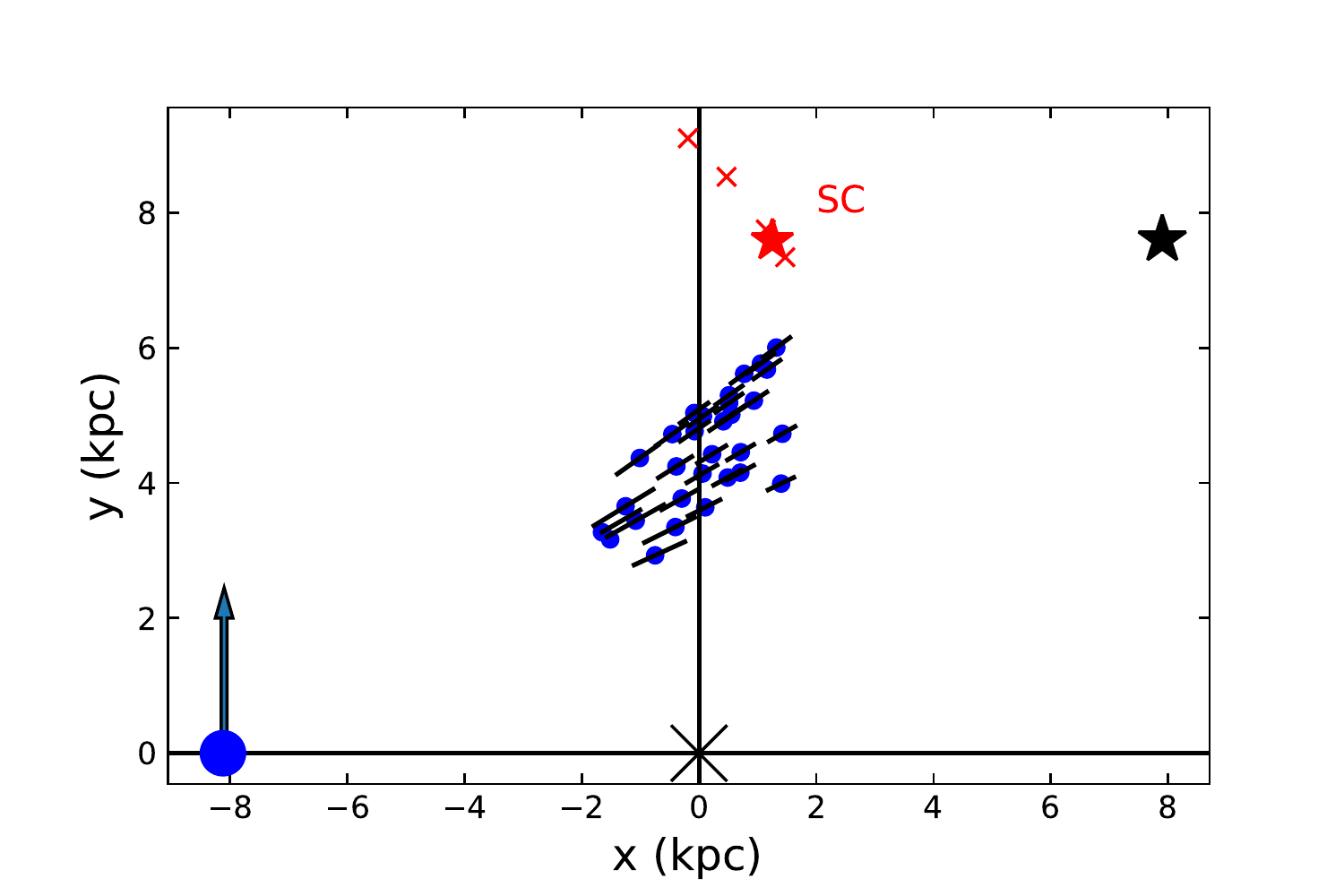}{0.45\textwidth}{(c)}}
\caption{
Location of regions where the leading clouds appear to be interacting with disk gas {  (small blue circles)}.
These were derived for the PM rotation curve with error bars showing the effect of  a $\pm5 \ \kms$ change in \vgsr. 
Red star and crosses are components of the Smith Cloud (SC) assuming a constant distance from the Sun for this object of 12.4 kpc.  
The large blue circle is at the position of the Sun {  and the "x" shows the Galactic center.}
{\it (a)} Projection of the cloud interaction locations on the x-z plane where x is in the direction of the Galactic Center (which is at 0,0), and z is distance from the Galactic plane.
{\it (b)} Projection of cloud interaction locations on the z-y plane, where y is perpendicular to the line of sight from the Sun to the Galactic Center. 
To some extent the correlated structure in this panel simply reflects the alignment of the Leading Component with the Smith Cloud in longitude and latitude.
{\it (c)} Cloud interaction locations in the x-y plane, which lies in the plane of the Milky Way.  
The black star shows the position of the Smith Cloud where it is projected to cross the Galactic plane in 27 Myr from the trajectory given in L08. 
 The sense of Galactic rotation is indicated by the arrow extending upwards from the symbol for the Sun.
}
\label{fig:interaction_locations}
\end{figure}

We assume that the velocity of material apparently interacting with a leading cloud is in fact disk gas in circular rotation and thus can be used to identify the location of the interaction within the Galaxy.
An identical line of argument was used for the Smith Cloud and produced a distance consistent with that derived from independent methods (L08).
While there are many uncertainties in this approach, and the distance to any individual cloud may have a large error,  this kinematic analysis has the potential to  provide information on the location of the leading cloud population  that is not otherwise available.
In respect of the uncertainties in derived distances we do not discuss the location of individual clouds but rather limit our comments to properties of the ensemble.
These do not vary greatly between the two rotation curves that were used.

 We analyze the cloud interactions with the two rotation curves described earlier:  a flat rotation curve with $V_{\theta}(R) = 230 \  \kms$, and the PM "Universal" rotation curve \citep{1996Persic,2019Mroz}. 
For objects at the longitudes of the leading clouds, a given positive value of \vlsr\ occurs at a single value of R, the distance from the Galactic Center, but at two distances from the Sun.
For the leading cloud sample the derived near distance is around 4 kpc, while the far distance is $>9$ kpc.
Given that the Smith Cloud is at a distance of 12.2 kpc we always choose the far kinematic distance for the leading clouds.
The velocity of the interacting gas is rarely determined with high precision: to estimate an effect of this uncertainly on our results we evaluated the kinematic distances at the \vlsr\ of the interaction $\pm 5 \ \kms$.
In the few cases where a velocity is greater than permitted by the Galactic rotation model the tangent point distance is adopted as the most likely.

Figure 12 shows derived distances to the interacting gas and by inference, to the leading clouds themselves, using the PM rotation curve  where the Sun is at $ x = -R_0 = -8.1$ kpc.  
The interaction sites of the leading clouds cluster around median values of $x,y, z = 0.2, 4.5, -0.6$ kpc, at a median distance from the Sun of 9.5 kpc.
This location is more than 8 kpc away from the spot where the Smith Cloud is expected  to cross the Galactic disk according to the trajectory of L08,  marked with a black star on panel (c) of \autoref{fig:interaction_locations}.

\begin{figure}
	\centering
	\includegraphics[width=0.41\textwidth]{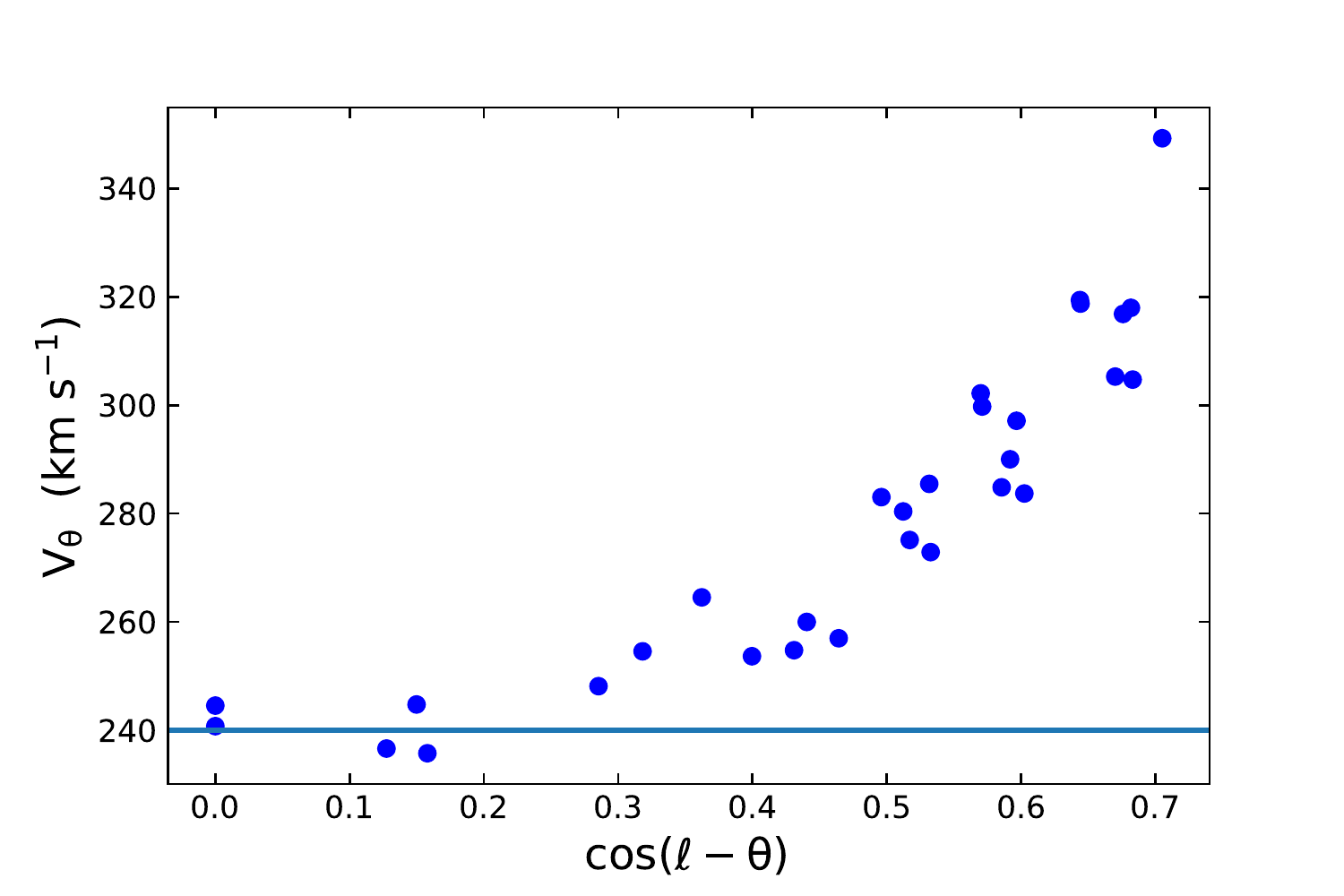}
	\caption{
		  Derived values of $V_{\theta}$ as a function of $cos(\ell - \theta)$ for $V_R = 0$, assuming that so close to the Galactic plane $V_z sin(b) = 0$. 
		  For very low values of 
		   $cos(\ell - \theta)$ any $V_R$ component  will be projected across the line of sight and  \vgsr\ will  be a function of  $V_{\theta}$ alone (\autoref{eq:VGSR}).
		   The horizontal line shows the adopted $V_{\theta} = 240$ \kms.
		  }
		\label{fig:Vtheta}
\end{figure}

\begin{figure}
	\centering
	\includegraphics[width=0.41\textwidth]{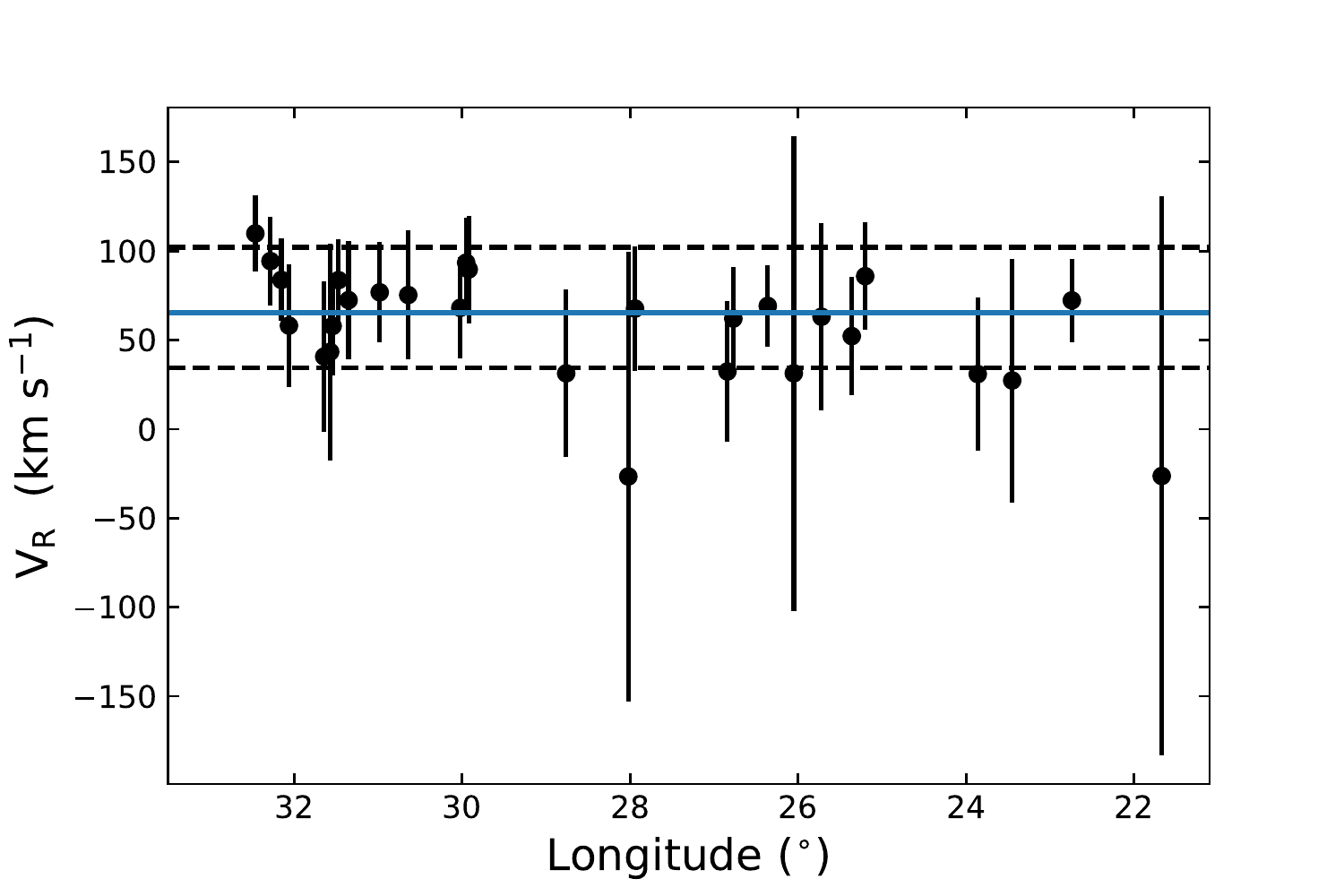}
	\caption{
		  Derived values of $V_R$ for $V_{\theta} = 240 \ \kms$ assuming that so close to the Galactic Plane $V_z  sin(b) = 0$.
		  The horizontal line shows the median value, 65 \kms.
		  Error bars show the effect of changing  \vtheta\  by $\pm 10$ \kms, and the dashed lines show the median values of $V_R$ that result.
		  }
		\label{fig:Long_VR}
\end{figure}

\subsection{Infall or Outflow?}
Throughout this work we have assumed that the Smith Cloud and the Leading Component are manifestations of  infall into the Milky Way.
There have been suggestions, however, that the Smith Cloud results  from a fountain process and is thus now moving away from the Galactic plane \citep{Sofue2004,Marasco2017}.
We believe that the properties of clouds in  the Leading Component resolve this issue.
For one, the Leading Component is found both above and below the Galactic plane at the same angle with respect to the plane as the Smith Cloud.  
This seems a very unlikely geometry for a supernova-driven superbubble originating in the Milky Way disk, but is unremarkable for an object passing through the Galactic plane.
Second, some leading clouds have components that connect them to  lower velocity emission, emission that is allowed by Galactic rotation (\autoref{fig:BV_maps}).
Were these clouds being accelerated outwards by a fast wind, the presumably stripped gas would be at a higher velocity.
These considerations lead to the same conclusion when applied to the Smith Cloud itself (\autoref{fig:BV_Smith}).
This discussion has no bearing on the origin of these objects,   but the case for their infall at the current time is persuasive.

\section{Space Velocity}
\label{sec:SpaceVelocity}

In a right-handed cylindrical coordinate system centered on the Galactic center where the angle $\theta$ is measured from the Sun-center line such that  $\ell \approx \theta $ at large distances, \vgsr\ can be written as the sum of three components, $V_{\theta}, \VR,$ and \vz 
\begin{equation}
\label{eq:VGSR}
 \vgsr =   [ - R_0 \ sin(\ell) \ V_{\theta}/R \ \ + V_R \ cos(\ell - \theta)]cos(b)  \ + V_z sin(b)  
 \end{equation}
where in this system the circular velocity $V_{\theta} < 0$.

\begin{figure}
	\centering
	\includegraphics[width=0.41\textwidth]{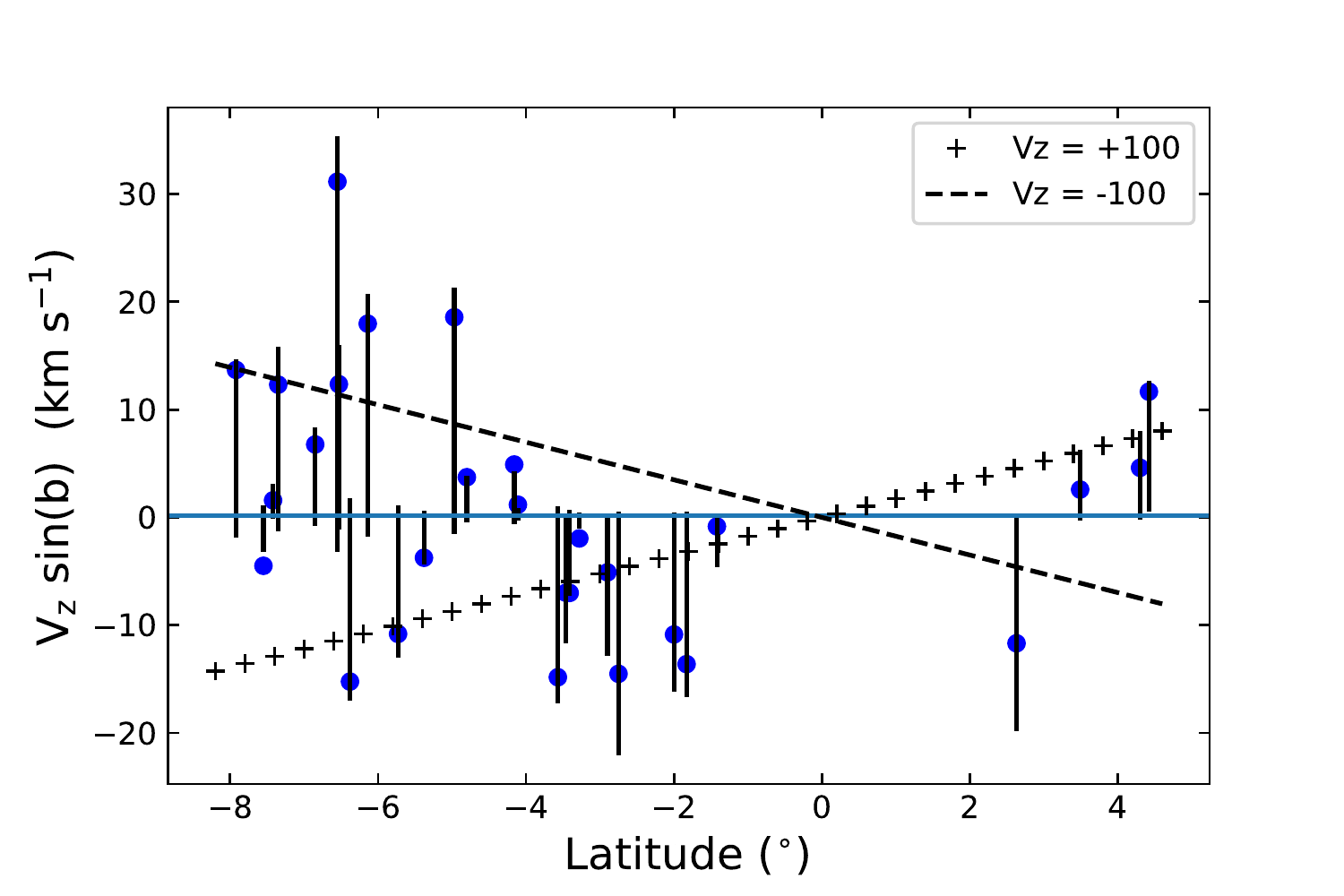}
	\caption{
		  Derived values of $V_z sin(b)$ for $V_{\theta} = 240 \ \kms$ and  $\VR = 65 \ \kms$. 
		  The horizontal line shows the median value, 0.2 \kms.
		  Error bars show the effect of changing the adopted \vtheta\  by  $\pm 10$ \kms.	
		  The lines formed with dashes and plus signs show values expected for $V_z = -100$ \kms\ and $V_z = +100$ \kms, respectively.
		  The small values of the derived $V_z sin(b)$ imply that $V_z$ cannot be determined accurately from these data.
		  }
		\label{fig:Lat_Vz}
\end{figure}

There are two limiting cases of interest:  first, near the Galactic Plane $sin(b) \approx 0$, so vertical velocities do not contribute significantly to \vgsr.
Second, at locations near the tangent points the quantity $cos(\ell - \theta) \approx  0$, and  the radial component of motion, $V_R$, does not contribute to \vgsr.
At these locations  \vgsr\ is a direct measure of $V_{\theta}$.
\autoref{fig:Vtheta} shows $V_{\theta}$ derived from these assumptions plotted against $cos(\ell - \theta)$.
The five clouds with the lowest values of $cos(\ell - \theta)$ give similar results for $V_{\theta}$ which we will adopt as its most likely value, 240 \kms,  indicated by a horizontal line in \autoref{fig:Vtheta}.

For a constant  $V_{\theta} = 240 \ \kms$, and, assuming that $V_z   sin(b)$  is small close to the Galactic plane, \autoref{eq:VGSR} can be solved for the  $V_R$ of the leading clouds.
The results are shown in \autoref{fig:Long_VR} where the median value, 65 \kms, is indicated by a horizontal line.
The error bars show the effect of a 10 \kms\ change in \vtheta.

In principle, having estimates for  \vtheta\ and  \VR\ should allow estimation of \vz\ for each cloud, but the determination is not robust, as sin(b) for these clouds is so small.  
\autoref{fig:Lat_Vz} shows \vz sin(b) vs. latitude with expectations for $V_z = \pm100$ \kms.
The small values and large uncertainties suggest that \vz\ cannot be determined reliably so close to the Galactic Plane.
The morphology and kinematics of the Smith Cloud indicate that its motion is to higher Galactic latitude, which requires $\vz > 0$.  
If the leading clouds share this motion they should have $\vz \ sin(b) < 0$ at $b < 0$ and lie along a curve like the one marked by the plus signs.

The estimated values for the space velocity of the Leading Component clouds are compared with these for the Smith Cloud from L08 in \autoref{tab:velocities}.
They are generally consistent.

The total space velocity of the Leading Component clouds is $V_{tot} > 250 \ \kms$  where the inequality indicates that we are unable to establish a reliable value for $V_z$.
 The Leading Clouds have a median line-of-sight velocity excess of 44 \kms\ with respect to the disk gas at their location.
This is in the direction of Galactic rotation.
As gas is stripped from these clouds (\autoref{fig:BV_maps}) it adds not only mass  but angular momentum to the disk.

\section{What is the Leading Component?}
\label{sec:discussion}
In this paper we have presented arguments for the existence of an organized interstellar structure whose orientation and kinematics are not consistent with Galactic rotation, but which has a connection to the Smith high velocity cloud.
The principle evidence for these claims is 
\begin{enumerate}
    \item A band of \hi\ emission crossing the Galactic plane a few degrees ahead of the Smith Cloud with an orientation  identical to that of the Smith Cloud.
 
    \item {  The clouds in this structure have velocities larger than allowed by Galactic rotation (\autoref{fig:Long_Vfor}) but the velocities are 
   virtually identical to the \vgsr\ of the  Smith Cloud,  indicating that there is a kinematic as well as a spatial connection between the structures.}

    \item Many of the leading clouds  are interacting with material at permitted velocities, confirming their peculiar kinematics (\autoref{fig:BV_maps} and  \autoref{fig:Long_interact}).
\end{enumerate}
 
We thus believe that the Leading Component is an anomalous feature of the ISM  related to the Smith Cloud, and despite its location in the Galactic plane this material had its origin as a  high velocity cloud.

Evidence of interaction with normal disk gas allows us to estimate that where the Leading Component crosses the Galactic plane near  longitude $25\arcdeg$ it is  9.5 kpc from the Sun and 4.5 kpc from the Galactic Center. 
This is a region of the Galaxy with significant star formation activity as indicated by the presence of numerous HII regions \citep{Anderson2012}, some of which have a parallactic distance identical to the median kinematic distance to the Leading Component \citep{Sato2014}.
The Leading Component has a neutral hydrogen mass comparable to that of the Smith Cloud, and 
it extends $\approx 17\arcdeg$ across the sky which, for our derived distance, makes its linear size $\sim 3$ kpc, identical to that of the Smith Cloud.

From estimates of interaction locations we derive the velocity components  for the leading clouds and find that they are similar to those of the Smith Cloud 
(\autoref{tab:velocities}). 
Because the two objects have a similar \vgsr\ (\autoref{fig:Long_VLSR_VGSR}) the similarity of the individual velocity components  is not surprising, but is nonetheless reassuring that we are observing a single phenomenon.
There is a gap between the Leading Component and the Smith Cloud of about $4\arcdeg$ in which there is no strong emission that can be associated with either object (see the left panel of \autoref{fig:long-lat_VGSR}).  
This would correspond to a gap of 0.6 - 1.6 kpc depending on the projection and assumed distance of the clouds.

The total space velocity of the Leading Component is $>250$ \kms, where the limit reflects our inability to measure $V_z$ accurately.
This is comparable to the L08 determination for the Smith Cloud of $V_{tot} = 296\pm 20$ \kms.
There is no information to suggest that either object has a velocity capable of escaping the Galaxy.
{  The estimates of cloud space velocity from \autoref{sec:SpaceVelocity} required   several approximations; a more complete study of the entire Smith Cloud and Leading Component taking into account the Galactic potential and dynamical effects is in progress.} 

The Smith Cloud has  been analyzed as an isolated system, perhaps the baryonic component of a dark matter subhalo \citep{Nichols2009,Nichols2014,Tepper-Garcia2018}, and the question of the survival of its gaseous component given its strong interaction with the Milky Way's circumgalactic medium has received considerable attention \citep{Gritton2014,Galyardt2016}.
The discovery of the Leading Component suggests that the Smith Cloud may not be an isolated object but  instead the more prominent manifestation of a system that stretches over more than 7 kpc on the sky with a gap between its two parts.
It is interesting that ``chains" of infalling gas occur naturally in some simulations that include simultaneous infall and outflow -- cloud chains that may survive to pass through a galaxy's disk \citep{Melso2019}.
Perhaps that is what we are observing here.
{  If the metallicity of the Leading Component is similar to that of the Smith Cloud at $\approx 0.5$ solar, it will lower the average metallicity of disk gas over the region of interaction. }

In \citet{Lockman2008} and subsequent publications there was speculation about the future of the Smith Cloud, as its trajectory indicated that it would enter and merge with the Milky Way disk in a few tens of Myr perhaps triggering a burst of star formation \citep{Alig2018}.
Given the existence of the Leading Component it appears that the merger is already underway.

\vskip 0.25in

Acknowledgments { The Green Bank Observatory is a facility of the National Science Foundation, operated by Associated Universities, Inc. 
The observations were made under GBT proposal codes 06C${\_}$038, 08A${\_}$014, and 09A${\_}$007.
We thank E.C. Kornacki for assistance with the data reduction {  and the anonymous referee for useful comments}.  
ECK,  NP, and CT were supported by the NSF REU programs at the National Radio Astronomy Observatory and the Green Bank Observatory. 
FJL acknowledges the influence of the classic work by \citet{Toomre1977} on the opening sentence of this paper.
}

\begin{deluxetable*}{lccccccccc}
\tablecolumns{10}
\tablecaption{Leading clouds detected in the GBT Observations. Columns as follow: (1) Galactic longitude; (2) Galactic latitude; (3) Maximum line brightness temperature; (4) LSR velocity; (5) Line width FWHM;  (6) Maximum $\hi$ column density; (7) observed angular size FWHM; (8) position angle with respect to the Galactic pole; (9) $\hi$ mass assuming a distance of 10 kpc. Columns (3)-(5) are from a Gaussian fit. Column (7) is from a 2D Gaussian fit to the channel map at the velocity of peak emission {  and is not corrected for beam smearing}. Column (9) is calculated assuming a distance from the Sun of 10  kpc. Uncertainties are propagated from the error associated with the Gaussian fits. 
If a cloud shows evidence that it is interacting with normal gas in the Milky Way disk the velocity of that gas is given as $V_{MW}$ in column (10).}
\label{tab:cloud_cat}
\tablehead{\colhead{$\ell$}  & 
\colhead{$b$} & 
\colhead{$T_\mathrm{pk}$} & 
\colhead{$\vlsr$} & 
\colhead{FWHM} & 
\colhead{$N_\mathrm{\hi}$} & 
\colhead{$\Theta_{obs}$} & 
\colhead{PA} &
\colhead{  $M_{\hi}\ $ (10 kpc)}  &
\colhead{$V_{MW}$}  \\
 \colhead{($\arcdeg$)} & \colhead{($\arcdeg$)} & \colhead{(K)} & \colhead{(\kms)} & \colhead{(\kms)} & \colhead{($10^{19}\ \cmm$)} & \colhead{($\arcmin$)} & \colhead{($\arcdeg$)} & \colhead{(\mo)} & \colhead{(\kms)} \\
\colhead{(1)} &  \colhead{(2)} & \colhead{(3)} & \colhead{(4)} & \colhead{(5)} & \colhead{(6)} & \colhead{(7)} & \colhead{(8)} & \colhead{(9)} & \colhead{(10)}
}
\startdata
	\hline\hline\noalign{\vspace{5pt}}	
			\noalign{\smallskip}
21.67 & +2.63 & 2.41$\pm$0.05 & 149.6$\pm$0.3 & 18.9$\pm$0.8  &  8.8 &   ---    &   ---   &  5.6e+03  & 119 \\ 
22.74 & +4.30 & 3.48$\pm$0.04 & 137.3$\pm$0.1 & 11.0$\pm$0.2  &  7.5 & 17.9$\times$15.0  &  125$\pm$6  &  1.2e+03 & 78 \\ 
23.45 & -2.00 & 1.84$\pm$0.04 & 146.2$\pm$0.1 & 12.3$\pm$0.3  &  4.4 &   ---    &   ---   &  2.4e+03  &  112 \\ 
23.86 & -3.57 & 1.83$\pm$0.03 & 136.6$\pm$0.2 & 19.7$\pm$0.6  &  7.0 & 22.6$\times$16.0  &  84$\pm$3  &  1.3e+03  & 103 \\ 
25.15 & -0.55 & 4.71$\pm$0.09 & 143.1$\pm$0.3 & 17.3$\pm$0.6  &  15.8 & 71.4$\times$57.0  &  27$\pm$6  &  1.0e+04 & --- \\ 
25.20 & +4.42 & 2.76$\pm$0.04 & 147.7$\pm$0.1 & 12.8$\pm$0.2  &  6.9 & 16.7$\times$14.0  &  134$\pm$6  &  1.4e+03  &  87 \\ 
25.36 & -3.41 & 2.63$\pm$0.05 & 132.2$\pm$0.2 & 16.9$\pm$0.5  &  8.6 & 72.0$\times$45.6  &  94$\pm$3  &  3.8e+03 & 91 \\ 
25.60 & -2.71 & 1.72$\pm$0.05 & 145.0$\pm$0.2 & 13.5$\pm$0.6  &  4.5 & 37.0$\times$13.6  &  168$\pm$1  &  1.1e+03 & 126 \\ 
25.72 & -1.42 & 1.70$\pm$0.10 & 146.7$\pm$0.3 & 9.1$\pm$0.9  &  3.0 & 17.8$\times$12.4  &  126$\pm$4  &  6.6e+02  & 105\\ 
25.74 & +4.78 & 1.06$\pm$0.03 & 142.6$\pm$0.3 & 24.8$\pm$0.8  &  5.1 & 17.2$\times$13.7  &  84$\pm$6  &  7.7e+02  & --- \\ 
26.05 & -2.90 & 2.13$\pm$0.06 & 140.8$\pm$0.2 & 13.3$\pm$0.5  &  5.5 & 41.3$\times$25.0  &  126$\pm$2  &  1.3e+03 & 114  \\ 
26.36 & +3.49 & 7.62$\pm$0.03 & 120.2$\pm$0.0 & 15.9$\pm$0.1  &  23.5 & 84.6$\times$45.6  &  1$\pm$2  &  1.1e+04  & 66 \\ 
26.77 & -3.28 & 3.16$\pm$0.03 & 127.1$\pm$0.1 & 15.6$\pm$0.2  &  9.6 &   ---    &   ---   &  6.9e+03 & 81 \\ 
26.84 & -2.24 & 1.44$\pm$0.06 & 136.8$\pm$0.4 & 12.4$\pm$1.0  &  3.4 & 36.3$\times$27.9  &  129$\pm$5  &  9.5e+02 & 116 \\ 
26.84 & -6.38 & 3.78$\pm$0.04 & 123.0$\pm$0.1 & 10.0$\pm$0.1  &  7.4 & 29.2$\times$14.7  &  132$\pm$1  &  1.9e+03 & 94 \\ 
27.94 & -4.11 & 2.90$\pm$0.04 & 132.7$\pm$0.1 & 12.5$\pm$0.2  &  7.0 & 47.9$\times$21.8  &  9$\pm$1  &  3.2e+03  & 87 \\ 
28.02 & -2.75 & 4.03$\pm$0.05 & 124.6$\pm$0.2 & 18.4$\pm$0.5  &  14.4 & 77.4$\times$37.8  &  127$\pm$1  &  8.6e+03  & 110 \\ 
28.76 & -1.83 & 11.35$\pm$0.08 & 121.8$\pm$0.1 & 12.8$\pm$0.1  &  28.2 & 37.4$\times$20.3  &  89$\pm$2  &  1.1e+04 & 96  \\ 
29.29 & -4.34 & 3.46$\pm$0.04 & 135.8$\pm$0.1 & 9.0$\pm$0.1  &  6.1 & 19.9$\times$15.3  &  124$\pm$7  &  1.0e+03 & --- \\ 
29.31 & -7.30 & 1.36$\pm$0.04 & 122.8$\pm$0.1 & 11.1$\pm$0.4  &  2.9 & 19.7$\times$11.0  &  51$\pm$2  &  4.3e+02 & --- \\ 
29.36 & -5.03 & 1.71$\pm$0.05 & 130.9$\pm$0.1 & 6.8$\pm$0.2  &  2.3 & 15.3$\times$11.8  &  163$\pm$3  &  2.8e+02  & --- \\ 
29.92 & -7.92 & 0.79$\pm$0.03 & 132.3$\pm$0.4 & 18.1$\pm$1.1  &  2.8 & 27.2$\times$18.0  &  16$\pm$4  &  1.1e+03 & 74  \\ 
29.95 & -6.14 & 3.49$\pm$0.03 & 128.2$\pm$0.1 & 13.5$\pm$0.2  &  9.1 & 22.0$\times$15.8  &  149$\pm$3  &  1.9e+03 & 63 \\ 
30.02 & -7.42 & 3.78$\pm$0.03 & 117.7$\pm$0.1 & 19.6$\pm$0.2  &  14.4 & 22.0$\times$16.8  &  98$\pm$3  &  1.3e+03 & 71 \\ 
30.07 & -3.92 & 5.30$\pm$0.11 & 121.3$\pm$0.2 & 15.2$\pm$0.6  &  15.6 &   ---    &   ---   &  1.7e+03  & --- \\ 
30.37 & -6.31 & 1.33$\pm$0.03 & 119.2$\pm$0.2 & 21.5$\pm$0.7  &  5.5 & 46.6$\times$38.0  &  91$\pm$7  &  2.0e+03  & --- \\ 
30.64 & -4.16 & 5.35$\pm$0.04 & 128.2$\pm$0.0 & 12.5$\pm$0.1  &  13.0 & 28.3$\times$15.4  &  137$\pm$2  &  2.5e+03 & 82 \\ 
30.98 & -6.85 & 1.01$\pm$0.02 & 119.2$\pm$0.5 & 29.6$\pm$1.2  &  5.8 &   ---    &   ---   &  2.3e+03 & 68 \\ 
31.35 & -4.80 & 2.40$\pm$0.03 & 121.7$\pm$0.1 & 15.2$\pm$0.3  &  7.1 & 54.6$\times$20.2  &  129$\pm$1  &  6.2e+02 & 76 \\ 
31.47 & -6.53 & 3.57$\pm$0.03 & 111.8$\pm$0.1 & 17.8$\pm$0.2  &  12.3 & 38.9$\times$22.1  &  162$\pm$1  &  3.1e+03 & 51 \\ 
31.54 & -7.55 & 3.86$\pm$0.03 & 105.2$\pm$0.1 & 19.4$\pm$0.2  &  14.5 & 32.1$\times$18.2  &  48$\pm$2  &  1.5e+03 & 65 \\ 
31.57 & -3.46 & 2.14$\pm$0.05 & 120.7$\pm$0.2 & 15.3$\pm$0.6  &  6.4 &   ---    &   ---   &  1.2e+03 & 95 \\ 
31.64 & -5.73 & 4.39$\pm$0.05 & 112.2$\pm$0.0 & 8.0$\pm$0.1  &  6.8 &   ---    &   ---   &  1.1e+03 & 85 \\ 
32.04 & -4.10 & 1.79$\pm$0.03 & 120.6$\pm$0.2 & 19.6$\pm$0.5  &  6.8 &   ---    &   ---   &  1.0e+03 & --- \\ 
32.06 & -5.38 & 5.22$\pm$0.03 & 112.9$\pm$0.1 & 15.2$\pm$0.1  &  15.4 &   ---    &   ---   &  3.2e+03 & 76 \\ 
32.15 & -7.35 & 2.73$\pm$0.03 & 110.2$\pm$0.2 & 18.8$\pm$0.4  &  10.0 & 30.0$\times$24.6  &  107$\pm$4  &  3.8e+03 & 50 \\ 
32.28 & -4.97 & 3.15$\pm$0.04 & 121.1$\pm$0.1 & 13.7$\pm$0.2  &  8.3 & 21.8$\times$17.1  &  78$\pm$7  &  1.2e+03 & 56 \\ 
32.46 & -6.55 & 2.46$\pm$0.03 & 123.4$\pm$0.1 & 14.0$\pm$0.2  &  6.7 & 44.1$\times$17.7  &  147$\pm$1  &  2.3e+03 & 43 \\ 
\noalign{\vspace{5pt}}
\enddata
\end{deluxetable*}

\begin{deluxetable*}{lccc}
\tablecolumns{4}
\tablecaption{Leading cloud population properties.
Distant-dependent quantities were calculated assuming a distance from the Sun, $d_{10} = 10$ kpc.  
Spectral properties were derived from a Gaussian fit at the location of the peak 21cm brightness.
The quantity \vgsr\ was calculated using a Galactic circular velocity of 230 \kms\ at the location of the Sun. 
\vdif\  is the difference in velocity between the cloud and the material it appears to be interacting with.
Angular sizes were derived from a Gaussian fit at the velocity of peak emission corrected by subtracting in quadrature the $10\arcmin$ effective angular resolution of the survey.
Column (4) gives the number of clouds used to establish each quantity. 
}
\label{tab:properties}
\tablehead{\colhead{Property}  & \colhead{Median} & \colhead{Range} &
\colhead{n clouds} \\
\colhead{(1)} &  \colhead{(2)} & \colhead{(3)}  & \colhead{(4)} \\
}
\startdata
	\hline\hline\noalign{\vspace{5pt}}	
			\noalign{\smallskip}
$T_{pk}  \ $ (K)         &  2.7   & 0.79 -- 11.35 & 38 \\
FWHM (\kms)             &  15.2  &  6.8 -- 29.6  & 38 \\
Peak \nhi\  $(10^{19}\ \cmm)$ &  7.0  & 2.3 -- 28.2 & 38 \\
\vgsr\  (\kms)            &  237   &  222 -- 248   & 38 \\
$\Theta_{maj} \ (\arcmin)$  &  28.3   &  11.6 -- 84 & 29 \\
$\Theta_{min} \ (\arcmin)$  &  15.0   &  4.6 -- 56 & 29 \\
PA $(\arcdeg)$  &          124  &   1 -- 168 &  29 \\
Diam  ($d_{10}$  pc)   &  57  &   25 -- 183 &  29 \\
$n \ \ (\cmmm \ \ d_{10}^{-1})$  &  0.43 & 0.13 -- 1.24 & 29 \\
$M_{\hi}\ (10^3 \ d_{10}^2 $ \mo)  & 1.6  & 0.28 -- 11.0 &  38 \\
$\vdif \ $ (\kms)    &  44  &   15 -- 81  & 30 \\
\noalign{\vspace{5pt}}
\enddata
\end{deluxetable*}

\begin{deluxetable*}{lccc}
\tablecolumns{4}
\tablecaption{Properties of three leading component clouds observed by the GBT at $10\arcmin$ resolution and by P2015 at $1\arcmin$ resolution. The P2015 diameters refer to the cloud core and were scaled to a distance of 10 kpc.
}
\label{tab:resolution_comp}
\tablehead{\colhead{Property}  & \colhead{G22.74+4.30} & \colhead{G25.20+4.42} &
\colhead{G26.84-6.38} \\
\colhead{} & \colhead{GBT, P2015} & \colhead{GBT, P2015} & \colhead{GBT, P2015} \\
\colhead{(1)} &  \colhead{(2)} & \colhead{(3)}  & \colhead{(4)} \\
}
\startdata
	\hline\hline\noalign{\vspace{5pt}}	
			\noalign{\smallskip}
Peak \nhi\   $(10^{20}\ \cmm)$ & 0.75, 1.7   & 0.69, 1.6 &  0.74, 2.2 \\
FWHM (\kms) \tablenotemark{a}  &  11.0, 5.8/24.3  &  12.8, 13.4  & 10.0, 3.0/14.5 \\
Diam  ($d_{10}$  pc)   &  37.5, 12.6  &  33.3, 13.3 &  50.0, 11.0 \\
$n \ \ (\cmmm \ \ d_{10}^{-1})$  &  0.65, 4.4 & 0.67, 3.9 & 0.48, 6.4 \\
PA $(\arcdeg)$  &   125, 122  &  134, 156 &  132, 128 \\
\noalign{\vspace{5pt}}
\enddata
\tablenotetext{a}{G22.74+4.30 and G26.84-6.38 have two spectral components in the P2015 data, one broad and one narrow.}
\end{deluxetable*}

\begin{deluxetable*}{lccc}
\tablecolumns{4}
\tablecaption{Space velocities of the Smith Cloud and the Leading Component.  
Smith Cloud entries are from the model in L08.  
Leading cloud entries are from the discussion in \autoref{sec:SpaceVelocity} where the the uncertainty in $V_R$ shows the effect of a 10 \kms\ change in the adopted $V_{\theta}$, and $V_z$ could not be determined.  }
\label{tab:velocities}
\tablehead{\colhead{Object}  & 
\colhead{$V_{\theta}$} & 
\colhead{$V_R$} &
\colhead{$V_z$} \\
\colhead{} & \colhead{\kms} & \colhead{\kms} & \colhead{\kms} \\
\colhead{(1)} &  \colhead{(2)} & \colhead{(3)}  & \colhead{(4)} \\
}
\startdata
	\hline\hline\noalign{\vspace{5pt}}	
			\noalign{\smallskip}
Smith Cloud & $270\pm21$   & $94\pm18$ &  $73\pm26$ \\
Leading Component  &  $240$  &  $65\pm35$ & --- \\
\noalign{\vspace{5pt}}
\enddata
\end{deluxetable*}

\facilities{GBT}

\bibliography{SmithCloudLeading_final}
\bibliographystyle{aasjournal}

\end{document}